# Sequential Federated Analysis of Early Outbreak Data Applied to Incubation Period Estimation

Simon Busch-Moreno[1] & Moritz U.G. Kraemer[1,2]

**Affiliations**

1. Department of Biology, University of Oxford, Oxford, UK

2. Pandemic Sciences Institute, University of Oxford, UK

**Corresponding author:** simon.buschmoreno@ biology.ox.ac.uk

**Abstract**

Early outbreak data analysis is critical for informing about their potential impact and interventions. However, data obtained early in outbreaks are often sensitive and subject to strict privacy restrictions. Thus, federated analysis, which implies decentralised collaborative analysis where no raw data sharing is required, emerged as an attractive paradigm to solve issues around data privacy and confidentiality. In the present study, we propose two approaches which require neither data sharing nor direct communication between devices/servers. The first approach approximates the joint posterior distributions via a multivariate normal distribution and uses this information to update prior distributions sequentially. The second approach uses summaries from parameters' posteriors obtained locally at different locations (sites) to perform a meta-analysis via a hierarchical model. We test these models on simulated and on real outbreak data to estimate the incubation period of multiple infectious diseases. Results indicate that both approaches can recover incubation period parameters accurately, but they present different inferential advantages. While the approximation approach permits to work with full posterior distributions, thus providing a better quantification of uncertainty; the meta-analysis approach allows for an explicit hierarchical structure, which can make some parameters more interpretable. We provide a framework for federated analysis of early outbreak data where the public health contexts are complex.

**Acknowledgements**

We thank members of PyMC Discourse who kindly answered our questions regarding federated analysis and posterior to prior updating methods. Many thanks to Jim Sheldon for his valuable input on system management. For the completion of this study all the following software packages were essential: for Bayesian sampling and statistics we use PyMC experimental (PyMC developers, 2024), PyMC (Abril-Pla et al., 2023), and ArviZ (Kumar et al., 2019), for numerical calculation support we use NumPy (Harris et al., 2020) and SciPy (Virtanen et al., 2020), for data wrangling we use pandas (The pandas development team, 2024), and for plots we use Matplotlib (Hunter, 2007).

**Authors Contributions**

Contributions to the study were as follows. S.B.-M.: conceptualisation, software, statistical methodology, analysis, writing. M.U.G.K.: conceptualization, administration, revision.

**Data and Code Availability**

All data and scripts are publicly available at:

https://doi.org/10.5281/zenodo.20804269

https://github.com/SimonErnesto/Sequential_Federated_Analysis

**Funding Sources**

M.U.G.K. acknowledges funding from The Rockefeller Foundation (PC-2022-POP-005), Health AI Programme from Google.org, the Oxford Martin School Programmes in Pandemic Genomics & Digital Pandemic Preparedness, European Union's Horizon Europe programme projects MOOD (#874850) and E4Warning (#101086640), Wellcome Trust grants 303666/Z/23/Z, 226052/Z/22/Z & 228186/Z/23/Z, the United Kingdom Research and Innovation (#APP8583), the Medical Research Foundation (MRF-RG-ICCH-2022-100069), UK International Development (301542-403), the Bill & Melinda Gates Foundation (INV-063472) and Novo Nordisk Foundation (NNF24OC0094346). The contents of this publication are the sole responsibility of the authors and do not necessarily reflect the views of the European Commission or the other funders.

**Declaration of interests**

None.

## 1. Introduction

Infectious disease outbreaks remain numerous and varied and continue to pose challenges to global health security. During the early phases of disease outbreaks and epidemics estimates of the key epidemiological parameters inform their likely impact and subsequent control strategies. This includes estimates of the reproduction number, incubation period, serial interval distribution, and disease severity. While there is general appreciation for the importance of these parameters, many challenges remain for estimating them early in epidemics. The primary obstacle preventing the joint estimation of critical epidemiological parameters is restrictions in data sharing, especially for data that include sensitive information about patients. For example, for estimating the incubation period, ideally the exposure data and symptom onset date are known at the patient level. Within countries, states or lower-level administrative units may act independently, restricting data sharing or device communication due to jurisdictional differences; between countries, few data sharing agreements exist ahead of outbreaks occurring, which severely restricts sharing and access (see Terry & Littler, 2024). Comparing estimates across contexts, however, remains important, especially when data at local levels are prohibitively small.

Federated analysis has emerged as an attractive paradigm for addressing challenges in data sharing. Federated learning is defined as an approach that allows multiple entities to collaboratively train a model without sharing their raw data which is achieved through local model updates being shared with a central server for aggregation. In a broad sense, a federated approach can refer to any form of decentralised data or analysis, from open/direct data sharing, going through data privatisation/anonymization techniques (e.g. cryptography), to strict federated analysis where no data but only derived results are shared (Rootes-Murdy et al., 2021). While anonymization techniques are intended to allow data sharing by distorting or encrypting the data, so it is not recoverable outside of a given analysis, such as differential privacy (e.g. Ju et al., 2022), federated analysis focuses on analysing data in situ at each provider's local device with only non-identifiable results being shared (Casaletto et al., 2023). For instance, federated computing techniques require no data sharing but often require direct communication between devices (i.e. machines, computing nodes, computing environment, etc.) or the sharing of re-constructible information (e.g. likelihoods), such as federated learning (e.g. Kidd et al., 2022). See Casaletto et al. (2023) for further examples of and discussion on these techniques.

Here we will focus on federated analysis, which attempts to technically overcome the challenges posed by privacy/confidentially restrictions. Federated analysis is an umbrella term describing different forms of decentralised data analysis aimed to respect confidentiality and privacy of data and devices (Rootes-Murdy et al., 2021; Casaletto et al., 2023). In a broad sense, a federated approach can refer to any form of decentralised data or analysis, from open/direct data sharing, going through data privatisation/anonymization techniques (e.g. cryptography), to strict federated analysis where no data but only derived results are shared (Rootes-Murdy et al., 2021). While anonymization techniques are intended to allow data sharing by distorting or encrypting the data, so it is not recoverable outside of a given analysis, such as differential privacy (e.g. Ju et al., 2022), federated analysis focuses on analysing data in situ at each provider's local device with only non-identifiable results being shared (Casaletto et al., 2023). For instance, federated computing techniques require no data sharing, but often require direct communication between devices (i.e. machines, computing nodes, computing environment, etc.) or the sharing of re-constructible information (e.g. likelihoods), such as federated learning (e.g. Kidd et al., 2022).

In many public health applications, we are faced with multiple obstacles, including difficulty in device access that would facilitate secure multi-party computation, orchestrated by a central server (e.g., conventional federated learning). Trust in anonymization techniques, which would facilitate sharing of non-identifiable data, by public health authorities also remains a challenge. A relevant example of a situation that requires a rapid analysis, but which may be approached with a relatively simple statistical model is the estimation of incubation periods early in disease outbreaks when countries individual datasets are small. Incubation period is defined as the time from infection to symptom onset (Kraemer et al., 2021). The incubation period is often used as a proxy for isolation policy following an exposure and thus has direct public health policy implications. For instance, recent Bayesian approaches (Virlogeux et al., 2016; Lauer, 2020; Miura et al., 2022; Madewell et al., 2023), show that simple models with well-known sampling distributions (e.g. Log-normal, Weibull or Gamma distributions) can efficiently estimate the incubation period of infectious diseases (see our supplementary material for further justification).

In this work we evaluate two methods for estimating the incubation periods early in outbreaks from multiple private (locally stored) data: i) Bayesian sequential updating and ii) a meta-analysis using simulated data of mpox and applying it to real datasets of influenza A virus

H7N9 and COVID-19. We apply both prior updating based on multivariate normal (*MvN)* approximations of the joint posterior distributions and *Meta-analysis* via a Bayesian hierarchical model to segmented simulated Mpox censored data, to segmented real H7N9 censored data, and to segmented COVID-19 doubly censored data. We also sample each full un-segmented dataset directly (*Direct Sampling*) for comparison. We expect that MvN-approx. and Meta-analysis will provide good estimates, approaching Direct Sampling, without requiring data sharing. By only sharing posterior distributions or posterior summaries this work enables more robust estimation of incubation periods in future outbreaks and provide a framework for estimating other epidemiological parameters.

Our method can be understood as a *sequential federated analysis* and as an alternative to classic federated learning methods such as secure multiparty computation, differential privacy, homomorphic encryption (see also Casaletto et al., 2023), Our proposed approach has no privacy-utility trade-off for cases which require low-level or mid-level complexity estimations —such as incubation periods— as it requires no data sharing and no device-to-device communication. While there are more complex computational frameworks and infrastructures that would enable multiparty computation and federated learning, they remain challenging to implement in current public health contexts, especially during early phases of disease outbreaks.

## 2. Methods

### *2.1. Application to Simulated Mpox Censored Data*

To simulate Mpox incubation periods we sample randomly from a Gamma distribution: $y_{i,n} \sim Gamma(\alpha, \beta)$, where shape $\alpha = \mu^2/\sigma^2$ and rate $\beta = \mu/\sigma^2$, with $\mu = 8$ and $\sigma = 3$. Then we produce censored data in the following way:

$$[y_l, y_u] = \begin{cases} [y_{i,n}, y_{i,n}] & with\, probability\; p \\ [max(0, y_{i,n} - L_{i,n}), y_{i,n} + U_{i,n}\,] & with\, probability\; 1 - p \end{cases}$$

Where $L_{i,n}, U_{i,n} \sim Uniform(1,5)$ and $p = 0.1$, with the additional constraint $U_{i,n} \geq L_{i,n} + 0.1$. The final reported interval is then transformed as $[y_l, y_u] = [y_l + 1, y_u + 1]$ to avoid values too close to zero. To emulate sample sizes relatively close to those typically found in real observations during emergency situations (e.g. Lauer et al., 2020), we simulated $n \ldots N =$ 6 sites each containing $i \dots I$ patients, where $I$ is randomly chosen from a range = [18,42]. This resulted in 156 datapoints, 6 sites with sample sizes = [37, 20, 23, 23, 24, 29].

We sample the resulting simulated data $y_l, y_u$ via the following Bayesian hierarchical model:

***Model 1(Direct Sampling)***

$$\alpha_s \sim HN(0.5)$$

$$\beta_s \sim HN(0.5)$$

$$\alpha_z \sim Normal(0, 0.1), n \dots N$$

$$\beta_z \sim Normal(0, 0.1), n \dots N$$

$$\alpha_n = \exp(\ln(7) + \alpha_s \alpha_z)$$

$$\beta_n = \exp(\ln(0.9) + \beta_s \beta_z)$$

$$\hat{y}_i \sim \begin{cases} Gamma(\alpha_n, \beta_n) & if\ y_l = y_u \\ ICG(\alpha_n, \beta_n) & if\ y_l \neq y_u \end{cases}$$

Where $\hat{y}_i$ is the likelihood over $i^{th}$ observations, a Gamma distribution when intervals are equal, otherwise an interval-censored Gamma distribution:

$$ICG = \ln(F(y_u|\alpha, \beta) - F(y_l|\alpha, \beta))$$

where $F$ is the cumulative density function (CDF) of the Gamma distribution (see Virlogeux et al., 2016), with shape $\alpha_n$ and rate $\beta_n$ use a non-centred parametrisation (see McElreath, 2020) with parameters varying over $n \dots N$ sites with locations 7 and 0.9 respectively, scales $\alpha_s$ and $\beta_s$ and offset distributions $\alpha_z$ and $\beta_z$. We chose locations parameters based on previous Mpox literature (e.g. Miura et al., 2022; Madewell et al., 2023), indicating Mpox incubation period means between 7-9 days and standard-deviations (SDs) between 3-5 days. We are aware this choice matches the very parameters we simulated, but we want to be consistent with the next applications of this model to real data, where literature-based priors are highly relevant. Priors for offset distributions have 0 mean and 0.1 SD, so when exponentiated they approach a standard Normal distribution. Half-normal ($HN$) scales are chosen to be 0.5 to provide more variability. This expresses the belief that our model should approach previously observed values (which we simulated to be as such) closely. Successful sampling should provide estimations which can recover the values of simulated parameters without overfitting.

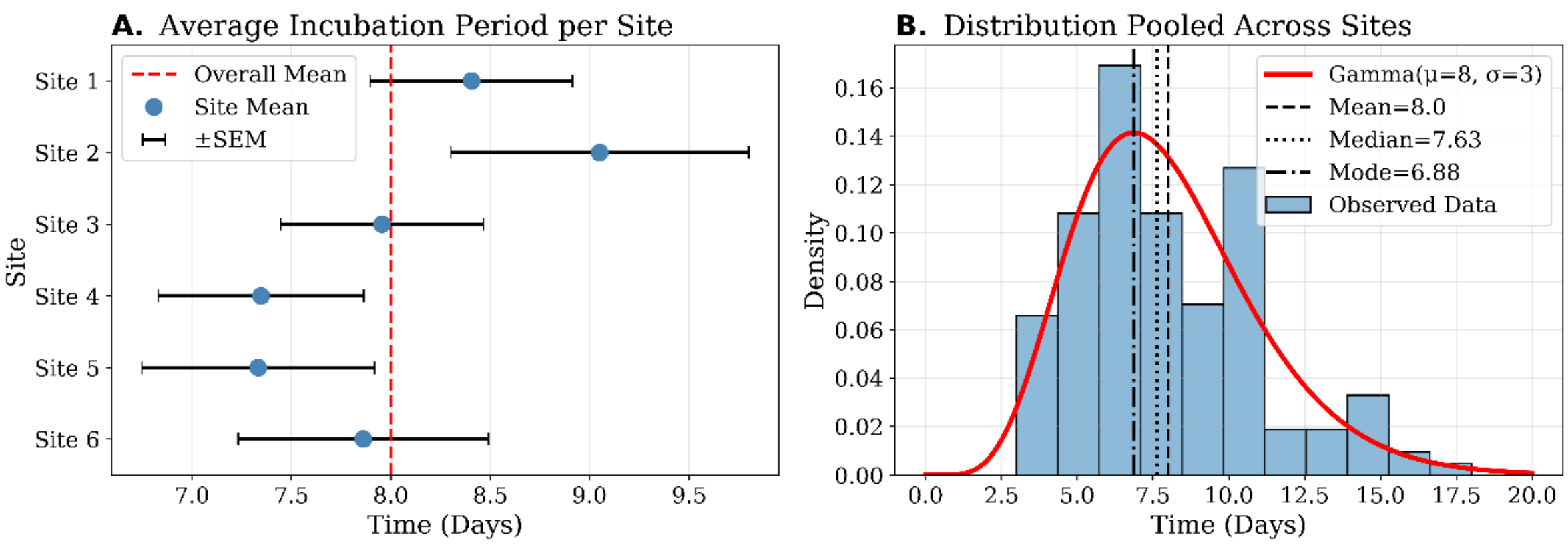

***Figure 1**. Simulated data summary. **A**: forest plot showing means and standard error of the mean (SEM) from each site. **B**: Averaged of pooled data, displaying gamma density and averages (mean, median, mode) of simulated Mpox incubation period.*

To test the multi-variate normal approximation (MvN-approx.) approach, we sample each site sequentially. The MvN-approx. approximates the priors for a model on site $n$ using the entire posterior distributions from a previously sampled site $(n-1)$. We sample the starting model (site 1) with a fixed parametrisation:

***Fixed Model (single-site)***

$$\alpha \sim TN_{0-10}(7, 1)$$

$$\beta \sim TN_{0-2}(0.9, 0.2)$$

$$\mu = \alpha/\beta$$

$$\sigma = \sqrt{\alpha/\beta^2}$$

$$\hat{y}_i \sim \begin{cases} Gamma(\alpha, \beta) & if\ y_l = y_u \\ ICG(\alpha, \beta) & if\ y_l \neq y_u \end{cases}$$

Where $\alpha$ and $\beta$ are assigned truncated-normal ($TN$) priors ranging from 0-10 and 0-2 respectively, and both having means 7 and 0.2 respectively (based on previous literature) and narrow SDs (1 and 0.2). Next, we construct the MvN approximation of $\mu$ and $\sigma$ for site $n$ using the priors from site $n-1$ and we repeat this iteratively until the last site. This process can be represented thusly:

***Model 2 (MvN-approx.)***

$$\bar{\theta}_{n-1}^{(p)} = p(\mu_{n-1}, \sigma_{n-1} | y)$$

$$\mu_n, \sigma_n \ = \ \bar{\theta}_{n-1}^{(p)} + \ L_{n-1}^{(p)} \cdot B$$

$$\alpha_n = \ \mu_n / \sigma_n$$

$$\beta_n = \ \mu_n / \sigma_n^2$$

$$\hat{y}_i \sim \begin{cases} Gamma(\alpha_n, \beta_n) & if\ y_l = y_u \\ ICG(\alpha_n, \beta_n) & if\ y_l \neq y_u \end{cases}$$

Where $\mu_n, \sigma_n$ are the new priors for current site $n$, $\bar{\theta}_{n-1}^{(p)}$ is the joint posterior mean taken from the joint posterior ($p$ = 1 … 2 posteriors from parameters: $\mu_{n-1}, \sigma_{n-1}$) obtained from previously sampled model $(n-1)$, $L_{n-1}^{(p)}$ is the Cholesky decomposition of the covariance matrix taken from the same joint posterior, and $B$ is a base normal distribution with standard deviation equal to one and mean equal to matrix of zeros with same size the joint posterior mean. Note that we use the PyMC-experimental implementation (PyMC developers, 2024), where $\bar{\theta}_{n-1}^{(p)} + L_{n-1}^{(p)} \cdot B$ corresponds to the non-centred parametrisation of an *MvN* distribution.

Finally, we use effect-sizes (posterior summaries) as input data for a hierarchical meta-analysis model. First, we obtain local effect-sizes by sampling each site independently with the *Fixed Model* presented before. Each site is sampled locally and independently with the fixed model, we use the resulting posteriors as effect-sizes $\delta_n = \alpha_n / \beta_n$ and errors $\epsilon_n = SD(\delta_n)$ as input (observed) data for a canonical Bayesian hierarchical meta-analysis model (e.g. Harrer et al., 2021). Note that $\delta_n$ corresponds to the posterior of $\mu = \alpha / \beta$ from each site and $\epsilon_n$ to each posterior's standard-deviation.

***Model 3 (Meta-analysis)***

$$\mu \sim Normal(0, s)$$

$$\tau \sim HN(e)$$

$$\zeta_n \sim Normal(0, 1)$$

$$\theta_n = \mu + \tau \zeta_n$$

$$\sigma_n \sim IG(3, \epsilon_n)$$

$$\hat{y}_{\delta_n} = \ Normal(\theta_n, \sigma_n)$$

Where $\mu$ is the general mean (location) with mean = 0 and adjustable standard-deviation $s$, $\tau$ is the between-study heterogeneity (scale) parametrised as a half-normal ($HN$) distribution with adjustable scale $e$, $\zeta_n$ is an across-sites offset distribution parametrised as a standard Gaussian with mean=0 and SD=1, $\theta_n$ are the latent effect-sizes per site *n* (non-centred parametrisation), and $\sigma_n$ are the errors across sites parametrised as an inverse-gamma ($IG$) distribution with mean=3 and SD=$\epsilon_n$. Effect sizes $\delta_n$ are treated as observed data for the $\hat{y}_{\delta_n}$ likelihood, a Gaussian distribution with mean $\theta_n$ and SD $\sigma_n$ parameters. How to adjust $s$ and $e$ fixed priors, depends on how much influence we want priors to have parameters' spread and associated estimates, for Mpox simulations analysis we chose $s = 5$ and $e = 5$ to allow more variability as data is quite homogeneous and we want to avoid overfitting.

The application of hierarchical models to meta-analyses is widely documented (see Gelman et al., 2013; Harrer et al., 2021), with half-normal ($HN$) and inverse-gamma ($IG$) priors being common choices. More sophisticated models are possible, but for present purposes (as proof of concept) we have chosen a relatively generic meta-analysis model which can be applied to all datasets in present analyses with minimal ad-hoc adjustments. A more throughout process of prior selection and calibration will be required for actual application purposes (e.g. see: Williams et al., 2018; Wang et al., 2023).

*2.2. Application to Influenza A (H7N9)*

We apply all models/approaches described above to data from Virlogeux and colleagues (2016). These data correspond to Avian Influenza A (H7H9) infections' maximum and minimum exposure period to disease onset ranges and additional details of individual patients. To emulate a multi-site environment, we chunked the original 395 datapoints into 9 chunks (sites) with randomly generated sample sizes = [45, 31, 41, 49, 47, 45, 48, 41, 48]. In the original study (Virlogeux et al., 2016) incubation periods from two groups of patients corresponding to non-fatal and fatal cases (G1 and G2 respectively) were estimated. Here we adapt the previously introduced *Model 1* to vary hierarchically over these two groups rather than over site (as present H7N9 data is generated from a single site).

So, we parametrise $\alpha$ and $\beta$ as: $\alpha_g = \exp(\ln(3) + \alpha_s \alpha_z)$ and $\beta_g = \exp(\ln(0.9) + \beta_s \beta_z)$. Where $g$ corresponds to [G1, G2], and priors for $\alpha$'s location = 3 and $\beta$'s location = 0.9 are derived from previous literature, where H7N9 mean is usually observed to be between 3-4 days with SD around 1-2 days (e.g. Guo et al., 2018). We adjust priors accordingly for the MvN-approx. model and the Meta-analysis model. For the latter we chose same values as for Mpox simulations, to allow more variability, as sites were artificially generated by chunking data. We do not use varying priors across sites for Direct Sampling, as the original data was generated from a single site, this could cause the model to overfit.

*2.3. Application to Corona Virus Disease 2019 (COVID-19)*

As a final test, we apply all approaches to COVID-19 data collected during early pandemic. This is also public data used for estimating COVID-19 incubation period by Lauer and colleagues (2020). Data corresponds to exposure intervals and symptoms intervals, analysed via a doubly-censored interval approach (Lauer et al., 2020). For our present purposes, analysis via our three approaches is intended as a proof of concept, so we do not use cases without both symptoms' intervals, keeping 172 cases as opposed to the 181 cases used in the original study. This leaves naturally chunked data, as cases are reported from different countries. We use each country as a singular Site, resulting in 23 sites with sample sizes: [84, 16, 13, 10, 8, 7, 6, 5, 3, 3, 2, 2, 2, 2, 1, 1, 1, 1, 1, 1, 1, 1, 1]. In general, Bayesian update is not influenced by sampling order (Kruschke, 2015), though in the present case there may be an association between site and sample size, because updates are not consistently generated from a single datum. Instead, each posterior $(n-1)$ used to updated prior $n$ comes from a disparate amount of data, which can range from 1 to 84 patients. So, just for this dataset, we ordered sites from higher to lower sample-size, as having the initial posterior generated from a larger amount of data may create initial priors which are more representative of overall data (for a similar approach to derive more informative priors and prior predictive checks see Kruschke, 2021).

The present COVID-19 data presents two challenges to sequential federated analysis approaches. Firstly, many sites will have a singular datapoint (i.e. a single case) which can serve to illustrate an advantage of Bayesian statistics, where priors make possible to sample models with a single datum (or without a datapoint at all). Secondly, doubly-censored data may be harder to sample, testing the capacity of the model to be accurate and capture uncertainty given low sample-sizes for a more complex model. Note that doubly-censored data implies two ranges of possible values, where lower and upper edges of the interval are not a single value each. Instead, we have a lower end with a left-right interval and an upper end with a left-right interval: [*lower[left,right], upper[left,right]*]. It is possible, then, to define a censored likelihood where lower-left ranges to upper-left and lower-right ranges to upper-right, with *ICG* sampling if lower-left≠upper-left or if lower-right≠upper-right and direct Gamma sampling otherwise. Thus, we simply replace the likelihood of previous models for the following likelihood for doubly-censored data:

$$\hat{y}_i \sim \begin{cases} Gamma(\alpha_n, \beta_n)_{left} & if\ y_l^{left} = y_u^{left} \\ Gamma(\alpha_n, \beta_n)_{right} & if\ y_l^{right} = y_u^{right} \\ ICG(\alpha_n, \beta_n)_{left} & if\ y_l^{left} \neq y_u^{left} \\ ICG(\alpha_n, \beta_n)_{right} & if\ y_l^{right} \neq y_u^{right} \end{cases}$$

Where *left* superscripts indicate the lower-left to upper-left interval edges, and *right* superscripts indicate the lower-right to the upper-right interval edges. These two intervals express the left-most range of possible lower to upper incubation period values and the right-most range of lower to upper incubation period values (see also: Yin et al., 2021). We also changed locations of $\alpha_n$ and $\beta_n$ to be consistent with previous literature, which shows means of COVID-19 to be around 6 days (e.g. Elias et al., 2021; Wu et al., 2022) with SDs around 3 days. Somewhat more conservatively, we choose a mean of 6 days and an SD of 3 days, which results in $\alpha_n = \exp(\ln(3) + \alpha_s \alpha_z)$ and $\beta_n = \exp(\ln(0.9) + \beta_s \beta_z)$. We adjust priors accordingly for the MvN-approx. model and the Meta-analysis model (for this analysis, we chose $s = 1$ and $e = 1$ as data is highly heterogenous and not constraining priors could elicit highly biased estimates).

## 3. Results

We used PyMC's (Abril-Pla et al., 2023) HMC sampler with 2000 tuning steps and 2000 samples with a tuning step of 0.95-0.99 to sample all models. Models sampled well with all effective sample sizes (ESS) > 2000, and $\hat{R} \cong 1$.

Similarity between distributions is formally compared via the overlap coefficient (Inman & Bradley, 1989): $OVL = \int_0^{\infty} \min[f(x; \alpha_1, \beta_1), f(x; \alpha_2, \beta_2)]dx$, where $\alpha_1$ and $\beta_1$ correspond to parameters from the first distribution, $\alpha_2$ and $\beta_2$ are the parameters from the second compared distribution, $x$ is a 30-day time-span, and $f$ is the Gamma PDF. We compared Direct Sampling with the MvN-approx. estimates ($OVL_{MvN}$) and with the Meta-analysis estimates ($OVL_{Meta}$) for all results. When OVL = 1, there is total overlap between distributions; if OVL = 0, there is no overlap between distributions.

### 3.1. *Simulated Mpox Results*

Models provide very similar general estimates, and they mainly differ in terms of uncertainty measures. Figure 2 summarises probability density functions (PDFs), cumulative density functions (CDFs) and posterior distributions $\mu$ from all three approaches applied to simulated Mpox incubation period data. OVLs are reported on panel A of Figure 2, both MvN-approx. and Meta-analysis approximate the Direct Sampling distribution closely. Table 1 summarises average measures from the PDFs, central tendencies have similar values across approaches, but Meta-analysis shows larger SDs as compared to the other two approaches.

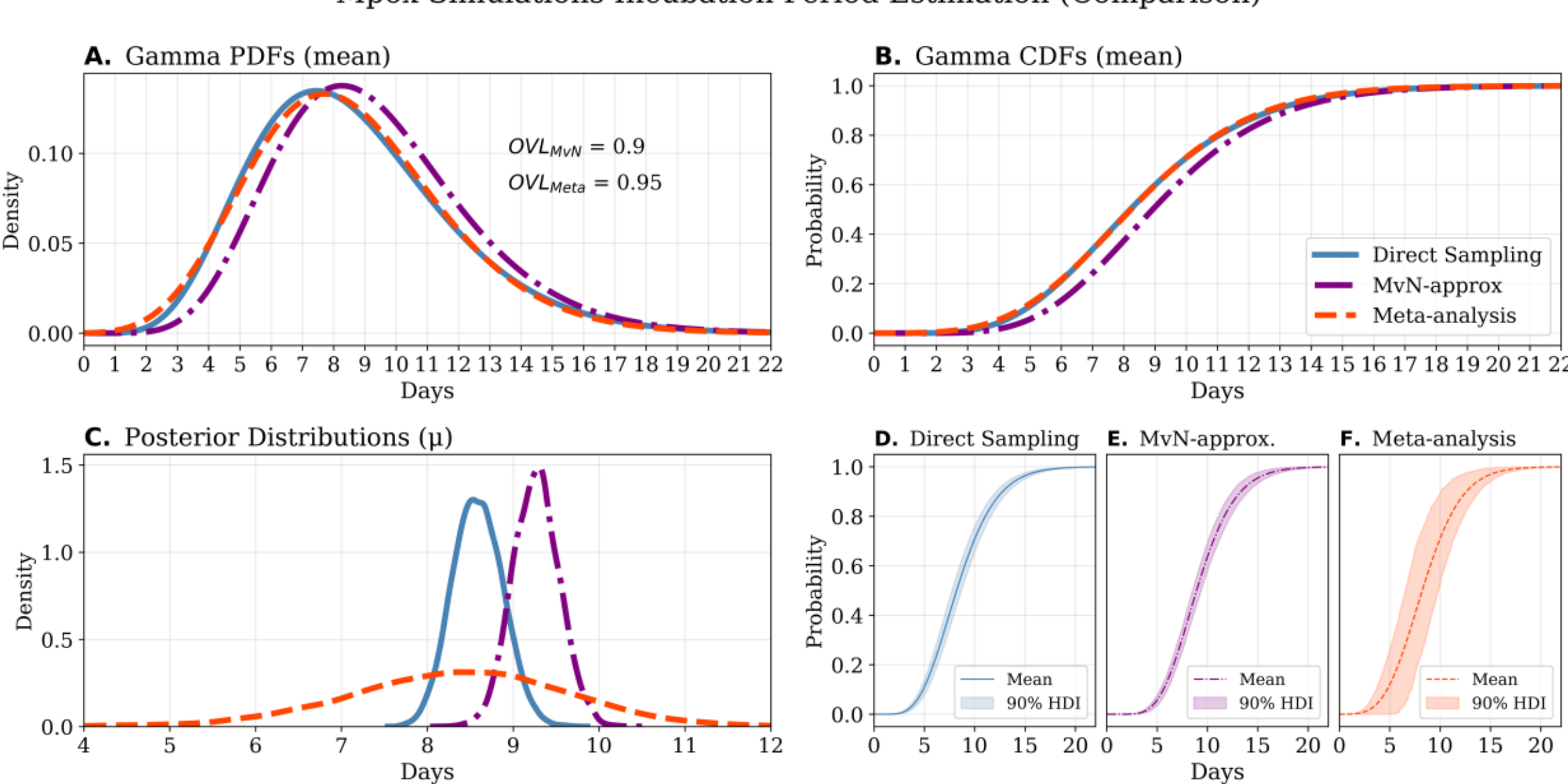

***Figure 2***. *Comparison between three approaches to simulated Mpox incubation period estimation.* ***A***: *Gamma probability density functions (PDFs) derived from posteriors of alpha and beta and averaged across samples, OVL correspond to overlap coefficients.* ***B***: *Gamma cumulative density functions from three approaches, derived from same parameters as PDFs.* ***C***: *Posterior distributions of Gamma mean mu.* ***D***: *Direct sampling CDF with 90% highest density intervals (HDIs).* ***D***: *MvN-approx. CDF with 90% HDIs.* ***F***: *Meta-analysis CDF with 90% HDIs.*

**Table 1**. Mpox Simulations Central tendency measures summary

| *Measure* | *Direct Sampling* | | | *MvN-approx.* | | | *Meta-analysis* | | |
|---|---|---|---|---|---|---|---|---|---|
| | **mean** | **SD** | **90% HDI** | **mean** | **SD** | **90% HDI** | **mean** | **SD** | **90% HDI** |
| **Mean** | 8.59 | 0.42 | [8.08, 9.19] | 9.27 | 0.27 | [8.82, 9.71] | 8.35 | 1.31 | [6.25, 10.45] |
| **Median** | 8.21 | 0.29 | [7.74, 8.69] | 8.93 | 0.26 | [8.50, 9.35] | 8.17 | 1.17 | [6.30, 10.11] |
| **Mode** | 7.45 | 0.29 | [6.98, 7.92] | 8.26 | 0.25 | [7.86, 8.67] | 7.49 | 1.3 | [5.36, 9.58] |

***Note***: *All averages, mean, median and mode, are calculated from* $f(x|\alpha, \beta)$, *where* $f$ *is the probability density function (PDF), and* $\alpha$ *and* $\beta$ *are the posterior distributions of Gamma shape and rate respectively.*

Table 2 summarises posterior distributions from all parameters, Direct Sampling tends to recover values closer to simulated parameters, the MvN-approx. recovers parameters with somewhat inflated values (with mean incubation $\mu$=9.27, more than a day larger than the 'true' value), and Meta-analysis recovers $\mu$ (8.35) well, but with inflated $\alpha$ and $\beta$ parameters. Means from a 30-day period fitted to the PDF $f(x|\alpha,\beta)$ indicate similar values, close to the 'true' 8-day mean incubation period, and with 90% highest-density intervals (HDIs) ranging from ~4 to ~13 days, close to simulated values (max range ~2-19 days), namely the models are 90% confident that the accurate incubation period of simulated Mpox falls between 4 to 13 days.

**Table 2**. Mpox Simulations Estimated parameters summary

| *Parameter* | *Direct Sampling* | | | *MvN-approx.* | | | *Meta-analysis* | | |
|---|---|---|---|---|---|---|---|---|---|
| | **mean** | **SD** | **90% HDI** | **mean** | **SD** | **90% HDI** | **mean** | **SD** | **90% HDI** |
| $\boldsymbol{\alpha}$ | 7.55 | 0.34 | [6.98, 8.03] | 9.38 | 1.07 | [7.68, 11.09] | 9.04 | 2.93 | [4.27, 13.61] |
| $\boldsymbol{\beta}$ | 0.88 | 0.03 | [0.83, 0.91] | 1.01 | 0.12 | [0.80, 1.20] | 1.04 | 0.24 | [0.65, 1.44] |
| $\boldsymbol{\mu}$ | 8.59 | 0.42 | [8.08, 9.19] | 9.27 | 0.27 | [8.82, 9.71] | 8.35 | 1.31 | [6.25, 10.45] |
| $\boldsymbol{\sigma}$ | 3.13 | 0.1 | [3.00, 3.27] | 3.04 | 0.21 | [2.71, 3.38] | 2.9 | 0.02 | [2.88, 2.93] |
| $\boldsymbol{f(x\|\alpha,\beta)}$ | 8.63 | 3.14 | [3.66, 13.53] | 9.17 | 2.99 | [4.24, 13.70] | 8.63 | 2.83 | [3.90, 12.81] |

***Note***: *All parameters are whole posterior distributions, $f(x|\alpha,\beta)$ is the probability density function (PDF), where $\alpha$ and $\beta$ are the posterior distributions of Gamma shape and rate respectively.*

As expected, Direct Sampling with a hierarchical model across sites shows a better balance between accuracy and uncertainty. The MvN-approx. tends to over-estimate values and with low uncertainty (i.e. the model may be over-confident). Meta-analysis is accurate for some parameters and shows higher estimate uncertainty respect to the other two approaches. See our supplementary materials Figure S1 and S2 for additional details.

### 3.2. *H7N9 Results*

Panel A of Figure 3 shows that Meta-analysis to Direct sampling OVL is again higher than MvN-approx. to Direct Sampling OVL. As expected from a larger sample-size, uncertainty is lower for Meta-analysis. Uncertainty seems too low for Direct Sampling, which may indicate that less restrictive priors could be needed (i.e. given larger sample-size wider scales could better capture data heterogeneity).

All approaches show central tendency measures consistent with general averages reported in the original study (Virlogeux et al., 2016), where mean incubation periods are around 3.5 days. On average, Direct Sampling estimates 3.4 days, MvN-approx. 3.11 days, and Meta-analysis 3.51 days. Note the high consistency of MvN-approx with previous literature (Guo et al., 2018).

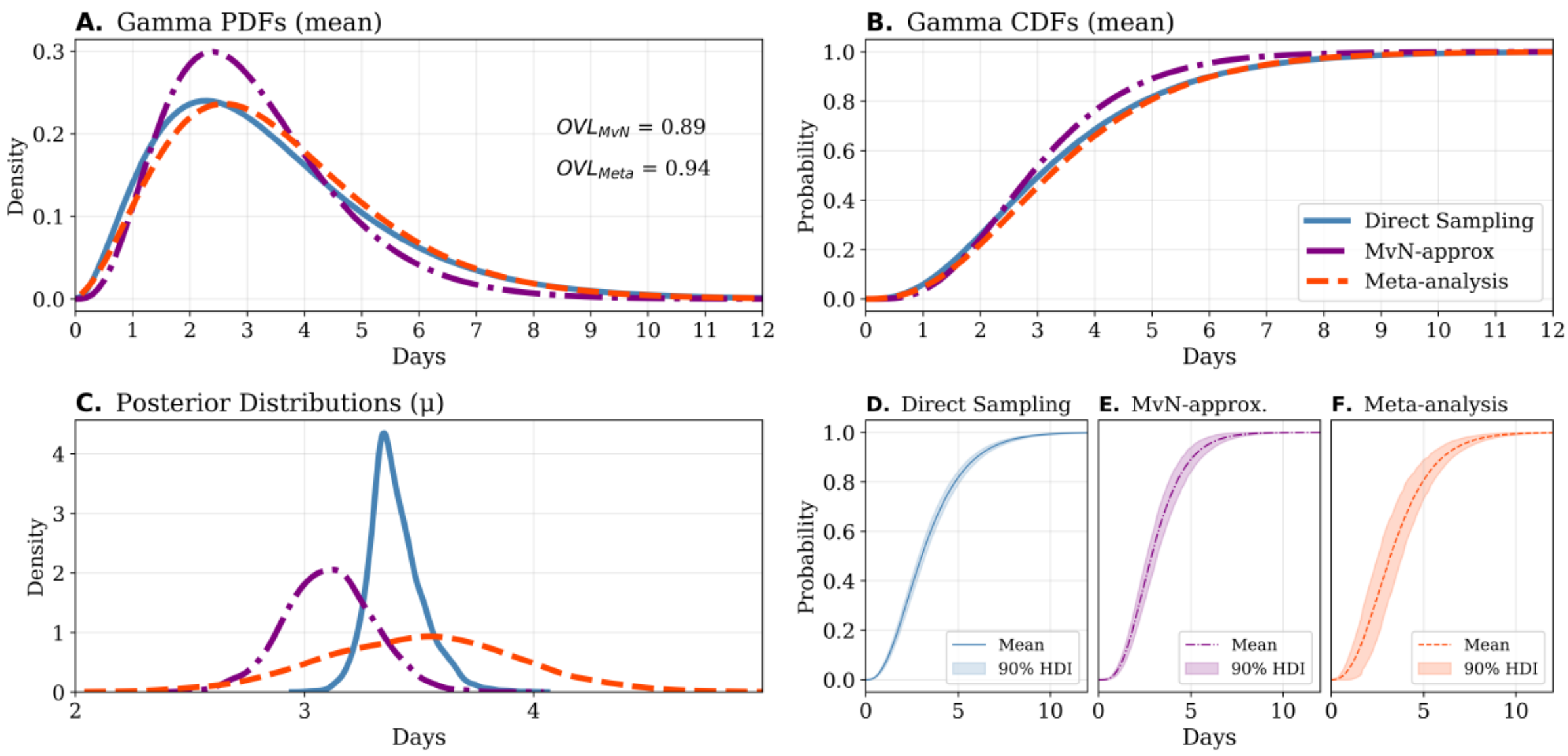

*Figure 3. Comparison between three approaches to Influenza A H7N9 incubation period estimation.* ***A****: Gamma probability density functions (PDFs) derived from posteriors of alpha and beta and averaged across samples, OVL correspond to overlap coefficients.* ***B****: Gamma cumulative density functions from three approaches, derived from same parameters as PDFs.* ***C****: Posterior distributions of Gamma mean mu.* ***D****: Direct sampling CDF with 90% highest density intervals (HDIs).* ***D****: MvN-approx. CDF with 90% HDIs.* ***F****: Meta-analysis CDF with 90% HDIs.*

As before, Meta-analysis shows SDs which 2-3 times larger than the other two approaches, but with closer HDI ranges in this analysis. This is also observed in Table 4, where means from a 30-day period fitted to the PDF $f(x|\alpha,\beta)$ retrieve similar values across approaches, and indicate plausible 90% HDI ranges of ~1-6 days, namely models indicate with 90% certainty that the accurate incubation period lays between 1 and 6 days. Both the MvN-approx. and Meta-analysis approaches seem to produce larger estimates for shape and rate parameters respect to Direct Sampling, though estimated $\mu$ and $\sigma$ have overlapping ranges (90% HDIs) across all three approaches.

**Table 3**. H7N9 Central tendency measures summary

| *Measure* | *Direct Sampling* | | | *MvN-approx.* | | | *Meta-analysis* | | |
|---|---|---|---|---|---|---|---|---|---|
| | **mean** | **SD** | **90% HDI** | **mean** | **SD** | **90% HDI** | **mean** | **SD** | **90% HDI** |
| **Mean** | 3.4 | 0.12 | [3.21, 3.58] | 3.11 | 0.19 | [2.80, 3.43] | 3.51 | 0.43 | [2.79, 4.20] |
| **Median** | 3.03 | 0.11 | [2.86, 3.22] | 2.88 | 0.18 | [2.60, 3.18] | 3.19 | 0.43 | [2.52, 3.94] |
| **Mode** | 2.28 | 0.1 | [2.12, 2.45] | 2.4 | 0.15 | [2.16, 2.66] | 2.52 | 0.53 | [1.65, 3.38] |

***Note****: All averages, mean, median and mode, are calculated from $f(x|\alpha,\beta)$, where $f$ is the probability density function (PDF), and $\alpha$ and $\beta$ are the posterior distributions of Gamma shape and rate respectively.*

**Table 4**. H7N9 Estimated parameters summary

| Parameter | Direct Sampling | | | MvN-approx. | | | Meta-analysis | | |
|---|---|---|---|---|---|---|---|---|---|
| | ***mean*** | ***SD*** | ***90% HDI*** | ***mean*** | ***SD*** | ***90% HDI*** | ***mean*** | ***SD*** | ***90% HDI*** |
| $\boldsymbol{\alpha}$ | 3.04 | 0.09 | [2.91, 3.19] | 4.44 | 0.44 | [3.75, 5.14] | 3.72 | 1 | [2.07, 5.25] |
| $\boldsymbol{\beta}$ | 0.9 | 0.02 | [0.86, 0.93] | 1.44 | 0.18 | [1.15, 1.72] | 1.05 | 0.2 | [0.73, 1.37] |
| $\boldsymbol{\mu}$ | 3.4 | 0.12 | [3.21, 3.58] | 3.11 | 0.19 | [2.80, 3.43] | 3.51 | 0.43 | [2.79, 4.20] |
| $\boldsymbol{\sigma}$ | 1.95 | 0.05 | [1.86, 2.03] | 1.48 | 0.13 | [1.28, 1.69] | 1.85 | 0.14 | [1.63, 2.09] |
| $\boldsymbol{f(x\|\alpha,\beta)}$ | 3.42 | 1.95 | [0.50, 6.18] | 3.05 | 1.45 | [0.77, 5.15] | 3.54 | 1.81 | [0.80, 6.16] |

***Note***: *All parameters are whole posterior distributions,* $f(x|\alpha,\beta)$ *is the probability density function (PDF), where* $\alpha$ *and* $\beta$ *are the posterior distributions of Gamma shape and rate respectively.*

Results from the three approaches are consistent with analysis results from the original study and are consistent between them. Although shape and rate ($\alpha$ and $\beta$) parameters are estimated to be somewhat higher by MvN-approx. and Meta-analysis approaches respect to Direct Sampling. General incubation period means and ranges are estimated in consistency with original analyses and previous literature. See supplementary figures S3 and S4 for extra details.

### 3.3. *COVID-19 Results*

Results provide central tendency measures (see Table 5) showing that all three approaches produce incubation period means between 5 and 7 days, with Direct Sampling and Meta-analysis showing more consistency with previous meta-analyses (Elias et al., 2021, Wu et al., 2022), while MvN-approx. shows values closer to the original study (Lauer et al., 2020). Again, the Meta-analysis approach shows higher uncertainty, with SDs 2-3 times larger than the other two approaches. Even so, as illustrated in Figure 4, all approaches render very similar Gamma distributions (note high OVLs in panel A).

**Table 5**. COVID-19 Central tendency measures summary

| *Measure* | *Direct Sampling* | | | *MvN-approx.* | | | *Meta-analysis* | | |
|---|---|---|---|---|---|---|---|---|---|
| | **mean** | **SD** | **90% HDI** | **mean** | **SD** | **90% HDI** | **mean** | **SD** | **90% HDI** |
| **Mean** | 6.19 | 0.27 | [5.69, 6.46] | 5.92 | 0.2 | [5.60, 6.25] | 6.28 | 0.65 | [5.27, 7.31] |
| **Median** | 5.69 | 0.16 | [5.47, 5.95] | 5.44 | 0.19 | [5.11, 5.74] | 6.12 | 0.61 | [5.22, 7.12] |
| **Mode** | 4.66 | 0.15 | [4.45, 4.90] | 4.44 | 0.21 | [4.11, 4.80] | 5.21 | 1.41 | [3.98, 6.63] |

***Note***: *All averages, mean, median and mode, are calculated from* $f(x|\alpha,\beta)$*, where* $f$ *is the probability density function (PDF), and* $\alpha$ *and* $\beta$ *are the posterior distributions of Gamma shape and rate respectively.*

Table 6 summarises parameters from the three approaches, indicating general coincidence in the estimation of shape, rate, mean and SD. Means from a 30-day period fitted to the PDF $f(x|\alpha,\beta)$ also show consistent ranges, in general agreement with both the original study and previous literature, with mean incubations around 6 days and 90% credible intervals falling between ~2 and ~11 days.

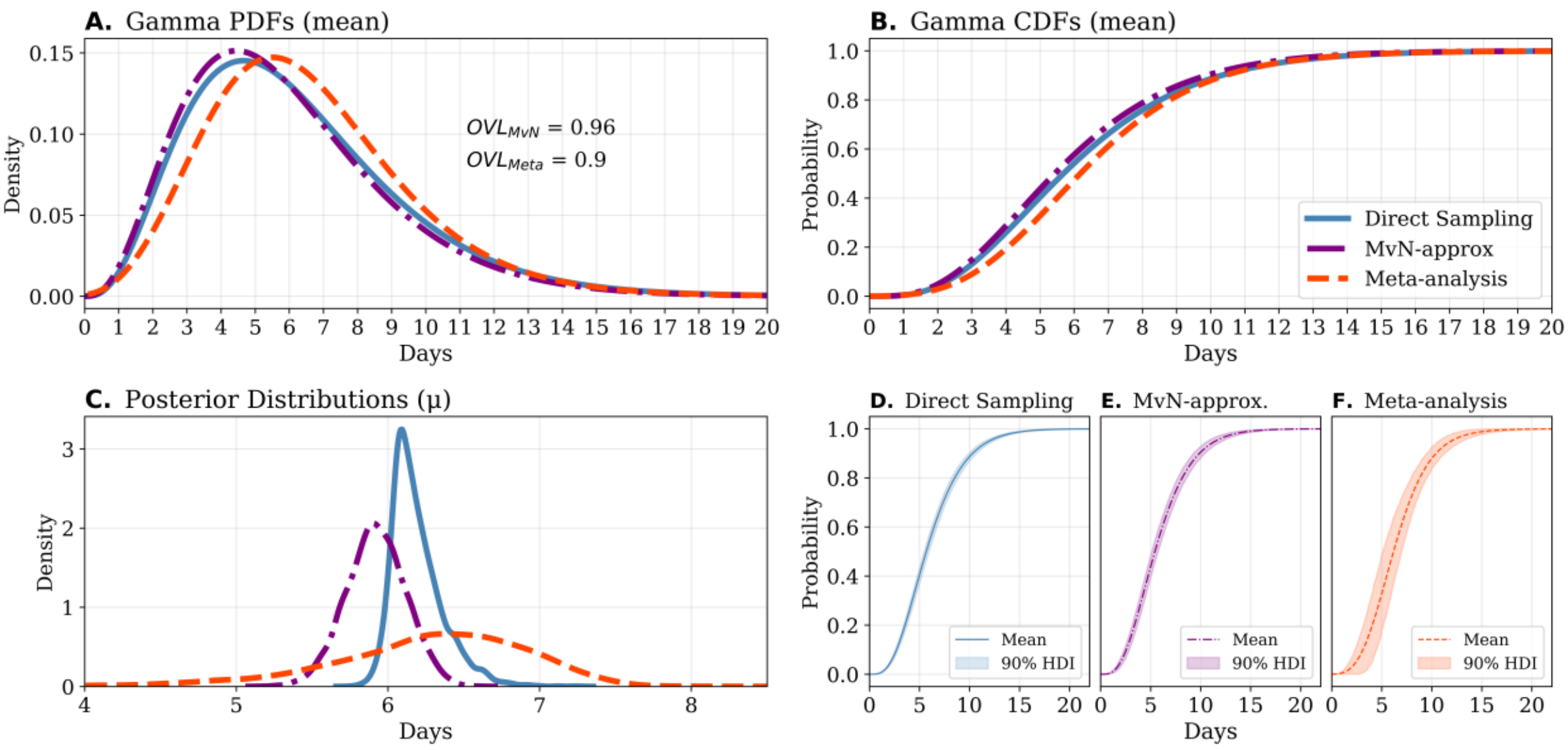

***Figure 4.*** *Comparison between three approaches to COVID-19 incubation period estimation.* ***A****: Gamma probability density functions (PDFs) derived from posteriors of alpha and beta and averaged across samples, OVL correspond to overlap coefficients.* ***B****: Gamma cumulative density functions from three approaches, derived from same parameters as PDFs.* ***C****: Posterior distributions of Gamma mean mu.* ***D****: Direct sampling CDF with 90% highest density intervals (HDIs).* ***D****: MvN-approx. CDF with 90% HDIs.* ***F****: Meta-analysis CDF with 90% HDIs.*

**Table 6**. COVID-19 Estimated parameters summary

| Parameter | Direct Sampling | | | MvN-approx. | | | Meta-analysis | | |
|---|---|---|---|---|---|---|---|---|---|
| | ***mean*** | ***SD*** | ***90% HDI*** | ***mean*** | ***SD*** | ***90% HDI*** | ***mean*** | ***SD*** | ***90% HDI*** |
| $\boldsymbol{\alpha}$ | 4.05 | 0.09 | [3.95, 4.18] | 4.03 | 0.36 | [3.44, 4.62] | 5.58 | 2.01 | [2.49, 8.83] |
| $\boldsymbol{\beta}$ | 0.65 | 0.01 | [0.63, 0.67] | 0.68 | 0.06 | [0.57, 0.78] | 0.84 | 0.26 | [0.43, 1.27] |
| $\boldsymbol{\mu}$ | 6.19 | 0.27 | [5.69, 6.46] | 5.92 | 0.2 | [5.60, 6.25] | 6.28 | 0.65 | [5.27, 7.31] |
| $\boldsymbol{\sigma}$ | 3.08 | 0.1 | [2.88, 3.18] | 2.96 | 0.15 | [2.72, 3.22] | 2.9 | 0.89 | [2.15, 4.23] |
| $\boldsymbol{f(x\|\alpha,\beta)}$ | 6.23 | 3.08 | [1.51, 10.78] | 5.84 | 2.92 | [1.33, 10.07] | 6.62 | 2.76 | [2.15, 10.60] |

***Note****: All parameters are whole posterior distributions,* $f(x|\alpha,\beta)$ *is the probability density function (PDF), where* $\alpha$ *and* $\beta$ *are the posterior distributions of Gamma shape and rate respectively.*

Models applied to COVID-19 data show similar estimates and produce similar incubation period distributions. Direct Sampling shows higher agreement with larger studies present in the literature, suggesting that some information may be lost during sequential sampling. This is to be expected, as present COVID-19 dataset is indeed sampled from multiple sites, namely the underlying data-generating process has multiple sources (see Figure S6 in the supplement). See supplementary Figure S5 for more details.

## 7. Discussion

We show that proposed approaches provide good alternatives for data analysis in a context where raw data cannot be shared nor devices can be accessed and orchestrated from a central instance. The first approach (MvN-approx.) approximates the sampling distributions parameters via a multivariate normal distribution (*MvN*). The *MvN* approximates the joint posterior from a previously sampled model. The second approach (Meta-analysis) consists of sampling each site separately and using posterior summaries of the sampling distribution mean as effect sizes for a subsequent meta-analysis. Both approaches were applied to simulated and real data, where models were run iteratively one site at a time. The rationale is that each approach, via a different process, should provide a good approximation of direct sampling of the whole dataset.

Both approaches were compared to a Direct Sampling approach, that is the sampling of the entire dataset via a hierarchical model. While the MvN-approx. approach can capture correlations at the joint posterior level, it cannot capture site-to-site variability, making its estimates more restrictive. Although it tends to produce a more stable structure driven by larger sample sizes. The Meta-analysis approach is agnostic to model structure and sampling procedure at the local level, as it can receive posterior summaries (effect-sizes and associated errors) produced by any type of model and sampling procedure, and it can provide hierarchical information from sites. However, it is more sensitive to prior selection due to low sample size (i.e. the number of sites) and model structure, where the number of datapoints is equivalent to the hierarchical priors' size (i.e. over-parametrisation). These advantages and limitations from both approaches were made clear by comparing them to direct sampling, which can capture both prior-level correlations and site-level hierarchical structures.

The low technical requirements of these approaches make them flexible to adapt to different situations, including but not limited to other incubation period models, causal models, generalised-linear models, and others. The only requirement each site (e.g. lab) has is to be able to sample a model locally and asynchronously share posterior distributions or posterior summaries via any generic data/text format (e.g. NetCDF, CSV, etc.). This facilitates analysis for contexts of strict restrictions to device-access and data-sharing. While more sophisticated implementations of federated analysis (for reviews see: Rootes-Murdy et al., 2021; Casaletto et al., 2023) rely on the relative relaxation of either shareability or access constraints, the present approach operates in a context where both are totally restricted. That is, when neither shareability nor access can be relaxed. Restrictions of this type can be detrimental for collaboration, restricting the ability to perform accurate inference and predictions for informing public health responses (Wartenberg & Thompson, 2010). Even though a technical solution is only part of a wider discussion for enabling open, shared, and trusted analyses early in outbreaks (e.g. Terry & Littler, 2024), we believe that our present approach can be a useful contribution for cases when a quick, practical, and robust solution is needed, which is often the case early in outbreaks.

For an initial response to outbreaks in a collaborative context, especially when cooperation between institutions at an international level is possible, the presented framework is simple and fast to implement without requiring centralisation of computing or data sharing. Here we presented two options for analysis of data at a local level but informed by the analysis previously carried on a different local dataset. We refer to each local dataset as a site, and the only requirement of present approaches is that there is an initial site sampling a full base-model and a final site sampling the last update of approximated parameters (MvN-approx.), or a final site compiling estimates from posterior summaries (asynchronously shared by other sites) and running a hierarchical model with such estimates (Meta-analysis). The sharing of summaries or distributions from one site to another can be asynchronous and does not require device communication protocols (e.g. port-to-port); or if communication is established this does not need to be on devices where data is stored. We acknowledge, however, that coordinating amongst groups might be challenging (though this may also apply to some standard approaches). Additionally, as summaries of parameters do not link to individual samples (e.g. one $\mu$ parameter instead of 21 likelihood parameters associated to 21 observation), the risk of reidentification and re-construction is much lower. Finally, any reportable information which may compromise privacy is only affecting the final site, which facilitates compliance according to regulations pertaining to that specific site.

There are several limitations for our framework: Firstly, their simplicity makes the applicability of more complex models limited, as models using latent variables, more complex processes (e.g. Gaussian processes), spatiotemporal structures, cannot be easily addressed via present approximations. More complex approaches are available, such as expectation propagation, which has the capacity to preserve information from each site (i.e. data chunk) by directly approximating local likelihoods based on previous approximations and priors (Vehtari et al., 2019). We have not directly explored this alternative approach here, but provided that a stable expectation propagation algorithm can be built by at least one of the sites and distributed to the others, and tests of these algorithms are passed, then it could be a promising extension to the framework presented here. Although, such an approach still requires caution, as sharing likelihoods or the totality of parameters could make this technique more susceptible to reconstruction or reidentification attacks (see Casaletto et al., 2023). Secondly, our approach is limited in that both MvN-approx. and Meta-analysis methods may perform better when models are standardised across sites. For example, flexibility on model implementation could induce erroneous pipelines when locally tailored models have extra parameters or lack parameters respect to other sites. To overcome this challenge Bayesian differential privacy approaches can be used to share data across sites (Ju et al., 2022) and thereby allowing one centralised analysis at a single site with full control over model implementation. Bayesian federated learning approaches centralise powerful flexible models which are run locally from a central server without data sharing (Kidd et al, 2023) but with direct communication between devices. Even the combination of federated learning and differential privacy has been proposed in such a way that noise is added to model parameters (Wei et al., 2020). Such approaches would involve strong software engineering capabilities across all participating sites.

Nevertheless, these approaches are promising alternatives to the one presented here (though further research is needed to identify sharing-access trade-offs), they remain non-compliant with the restriction (e.g. legal) framework we face in many situations during early outbreaks. Other alternatives, such as asynchronous (i.e. off-line) applications of federated learning (akin to expectation propagation), able to operate with strong access/communication restrictions, can be promising provided further research can demonstrate their compliance with severe restrictions.

It is important to emphasise, however, that decentralised approaches, such as federated analysis, are not a silver bullet for solving data privacy issues, but longer-term collaborative solutions to develop more ethical and secure data sharing systems are also required (Bak et al., 2024). We have presented an option of decentralised analysis which respects both privacy and access restrictions. Even so, we are aware that beyond the framework of early outbreak spontaneous collaboration, solutions outside the analytical technical domain are essential. Technical solutions to addressing early and coordinated disease outbreak analyses cannot be successful without the trust and collaboration between countries.

## 8. Conclusion

Our approaches show that it is possible to analyse data when data sharing and device access is restricted. When models do not strictly require a hierarchical structure, an MvN-approx. approach provides a reasonable way of obtaining quick and stable estimates via a prior update approach. Similarly, a meta-analysis approach can provide a model with hierarchical structure, which also allows sampling flexibility at each independent location, as meta-analysis can operate with effect-sizes produced by any type of model. We discussed the disadvantages of these two approaches, mainly focusing on their limited capacity to tackle more complex and flexible models. Alternative approaches such as expectation propagation could ameliorate this problem in the future, but more research is needed on their implementation in early outbreak settings.

**Supplement**

Sequential Federated Analysis of Early Outbreak Data Applied to Incubation Periods Estimation

Simon Busch-Moreno[1] & Moritz U.G. Kraemer[1,2]

**Affiliations**

1. Department of Biology, University of Oxford, Oxford, UK
2. Pandemic Sciences Institute, University of Oxford, UK

The present supplement includes further justification for the use of gamma sampling distributions and plots with additional summaries of distributions estimated from simulated Mpox data, publicly available Influenza A H7N9, and corona virus disease 2019 COVID-19. Plots include Gamma distributions sumamries with average parameter estimates, and forestplots containing site-level estimates summaries.

To further justify the choice of sampling distribution (Gamma), we emphasise that –theoretically speaking– a Gamma distribution models the sum of *k* i.i.d. exponential random variables, which corresponds to the waiting time until the *k*-th event in a Poisson process (for proof see: Ross, 2019). Lacking exact time counts to model events via a Poisson process, and assuming the incubation process as continuous in time, the Gamma distribution is a theoretically optimal choice. Although other distributions (e.g. log-normal or Weibull) can also provide appropriate results. However, for the sake of simplicity, as our focus is purely on the application of estimation methods for a sequential federated analysis approach, we will exclusively use Gamma sampling distributions. We note, nonetheless, that the methods we present here are agnostic to sampling distributions and in more detailed applications they could eventually incorporate and compare models using different sampling distributions (e.g. log-normal, Weibull, Gamma, etc.).

Presently, we will focus on sequential sampling methods that can preserve Hamiltonian Monte Carlo (HMC) sampling, a variant of Markov Chain Mote Carlo (MCMC) sampling. HMC usually provides more efficient sampling of high dimensional spaces, allowing for more diverse models, plus more informative convergence diagnostics for a better assessment of model sampling and posteriors (Betancourt, 2017). We will introduce two approaches: 1) Prior updates based on a multi-variate normal (MvN) approximation of the joint posterior distributions. This method requires sampling data locally at an initial site (e.g. lab, hospital, etc.) and sharing the posterior distribution of relevant parameters, which will be used to generate the MvN approximation to define the prior distributions at a new site, this new site follows the same procedure with a third site and so on and so forth (for a detailed explanation of Bayesian updates see Kruschke, 2015; McElreath, 2020). 2) Meta-analysis via a Bayesian hierarchical model. This method requires sharing posterior summaries, in particular containing parameters' posterior means and standard-deviations, which will be used as effect-sizes and errors within the meta-analysis model (for more details on Bayesian meta-analysis see Gelman et al., 2013; Harrer et al., 2021).

Note that site-level estimtes are all named effect-sizes for consistency, but they should be interpreted in different ways. Direct Sampling effect-sizes refer to the site means $\mu$ estimated from a whole data hierarchical model (i.e. varying parameters). Meta-analysis effect-sizes refer to the meta-analysis hierarchical model site-by-site estimated means $\mu$ ($\theta_n$ in the meta-anaysis hierarchical model). MvN-approx. effect-sizes refer to the $\mu$ estimates from a locally sampled fixed model, where models after the firstly smapled receive $\mu_{n-1}$ information to parametrise priors.

Note Figure S4 panel B, where MvN-approx. approaches better results for G1 and G2, closer to the original study, probably because artificial data chunking has biased the estimates of the other two approaches (without disregarding the possibility of biases present in the original study and our approach due to sampling via a fixed model). We do not expect this effect when data are naturally generated from different sources, which is evidenced by Figure S6, where effect sizes, though differently patterned, remain consistent across approaches (panels A to C), notwithstanding the difficulty of sampling from very low sample-size sites.

Another noteworthy pattern in present plots is the very low uncertainty expressed by Direct Sampling (see Figures S1, S3 and S5, Panels A), which may be due to highly constraining priors. In real situations, less informative priors may be required after prior sensitivity analysis. Oppositely, Meta-analysis hierarchical models induce high uncertainty (see Figures S1, S3 and S5, Panels B), probably due to being over-parametrised (i.e. a much higher number of parameters than datapoints). Even so, they manage to recover estimates with good accuracy.

## *S1. Additional Mpox Simulations Summaries*

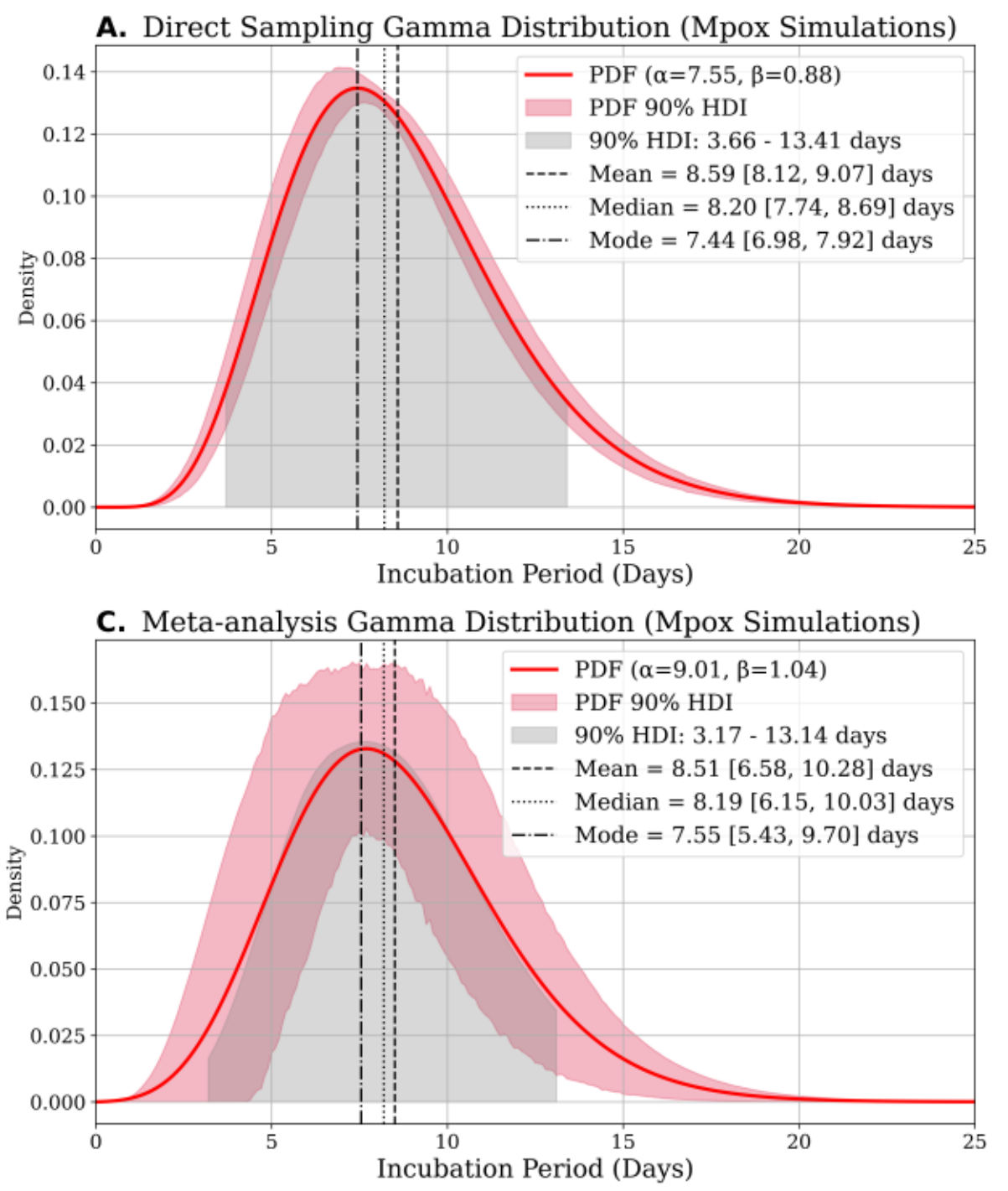

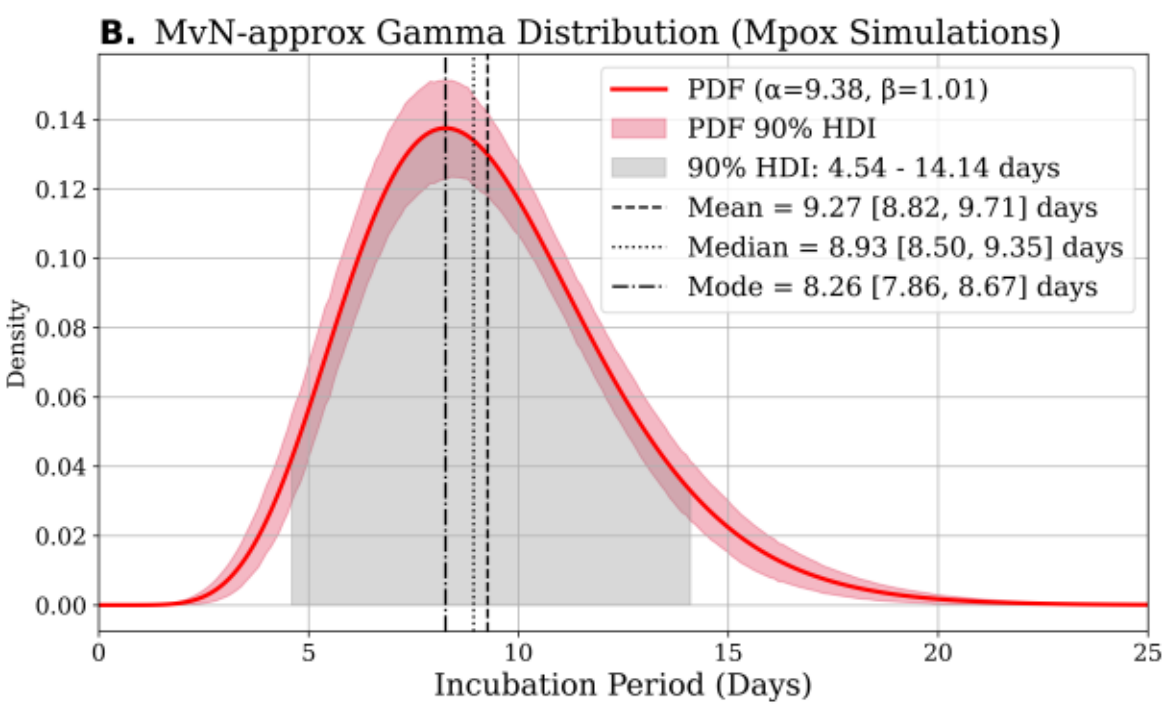

**Figure S5**. Mpox Simulations PDF summaries. Gamma distributions are computed as $f(x|\alpha,\beta)$ where $f$ is the PDF of the Gamma distribution and $\alpha$ and $\beta$ are shape and rate (their posterior means are shown in plots). The grey shadow indicates the 90% distribution highest density interval (HDI). Dashed, dotted and das-dotted line indicate mean, median and mode of the Gamma distribution respectively, their respective HDIs in brackets.

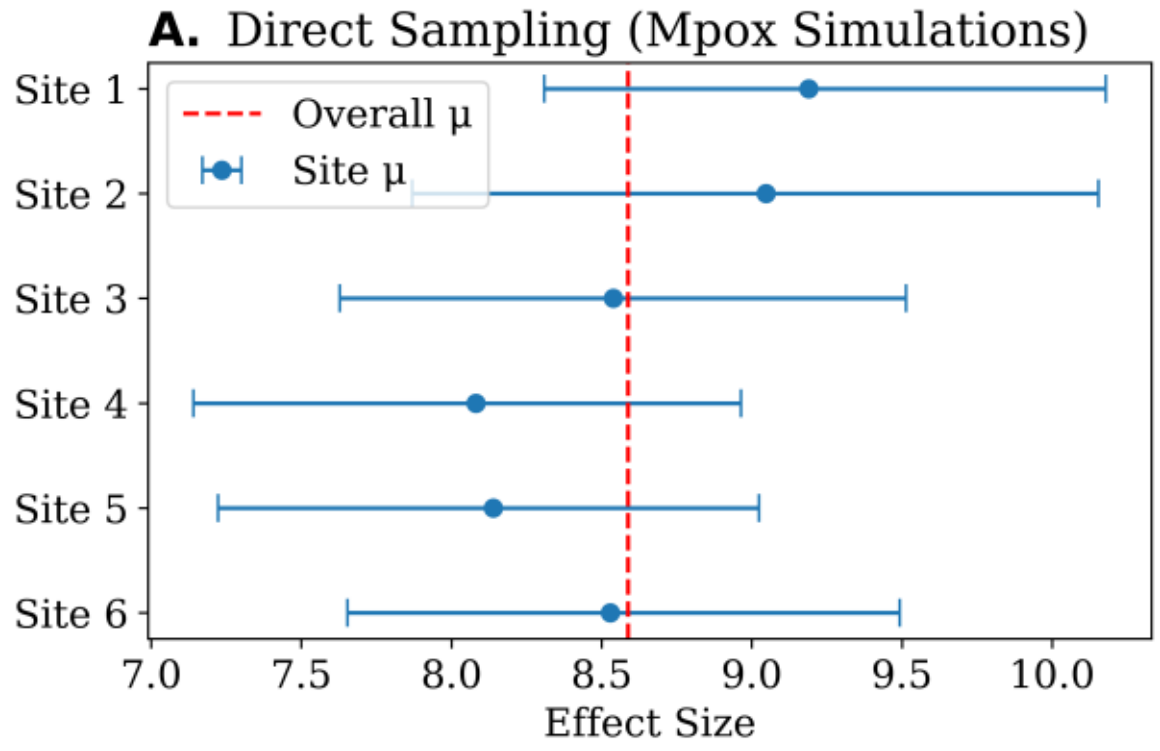

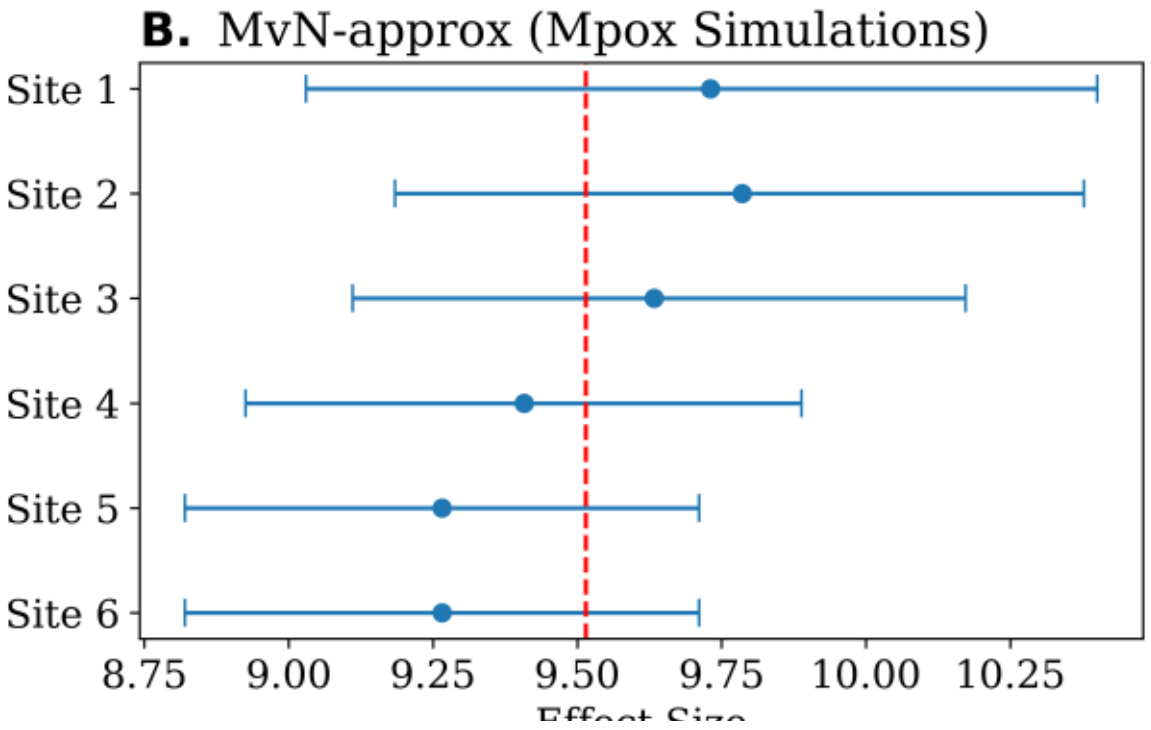

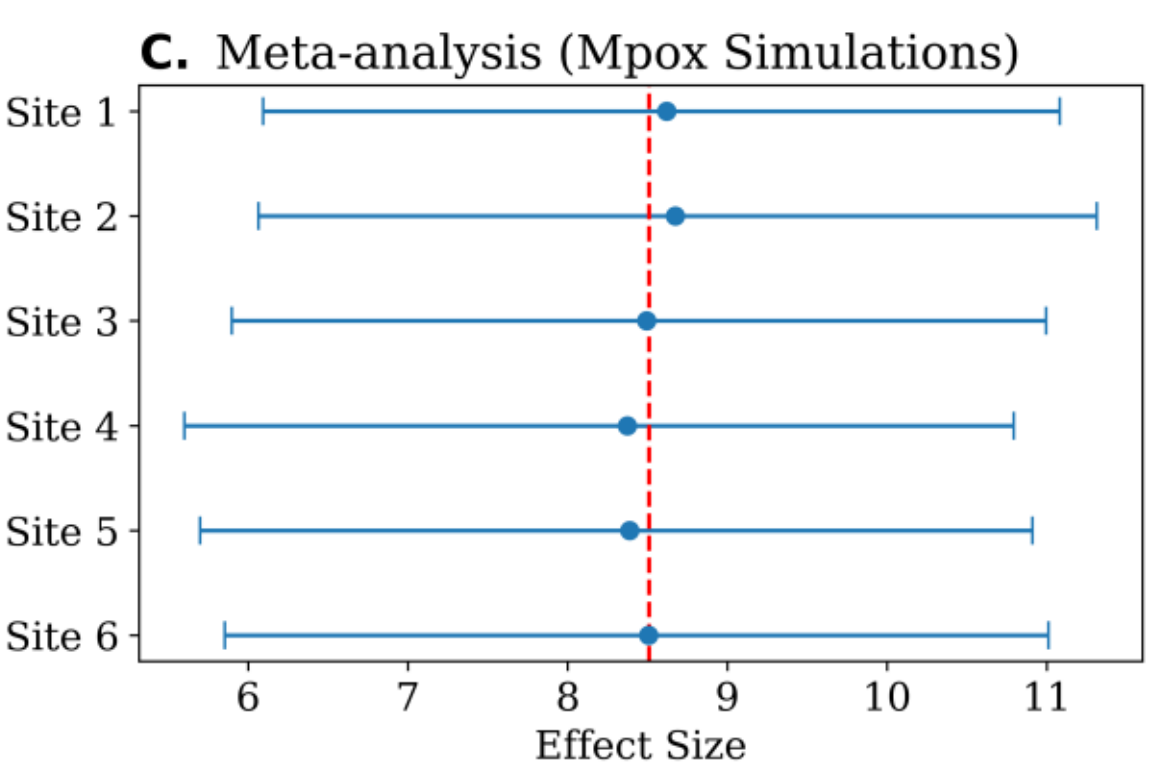

**Figure S6**. Mpox Simulations Gamma mean ($\mu$) summaries. **A**: Direct Sampling effect-sizes refer to the site means $\mu$ estimated from a whole data hierarchical model (i.e. varying parameters). **B**: MvN-approx. effect-sizes refer to the $\mu$ estimates from a locally sampled fixed model, where models after the firstly smapled receive $\mu_{n-1}$ information to parametrise priors. **C**: Meta-analysis effect-sizes refer to the meta-analysis hierarchical model site-by-site estimated means $\mu$ ($\theta_n$ in the meta-anaysis hierarchical model).

## *S2. Additional H7N9 Summaries*

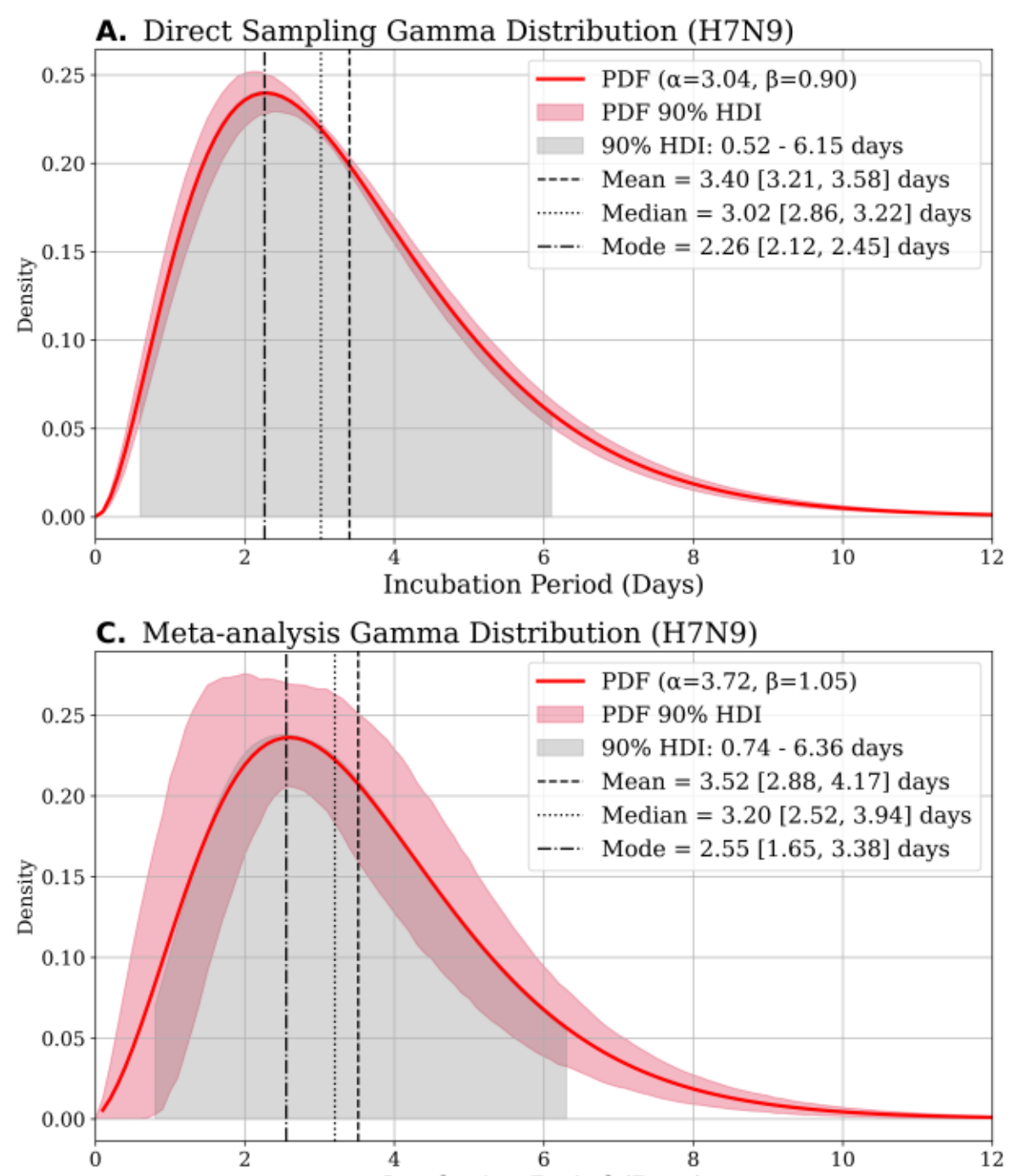

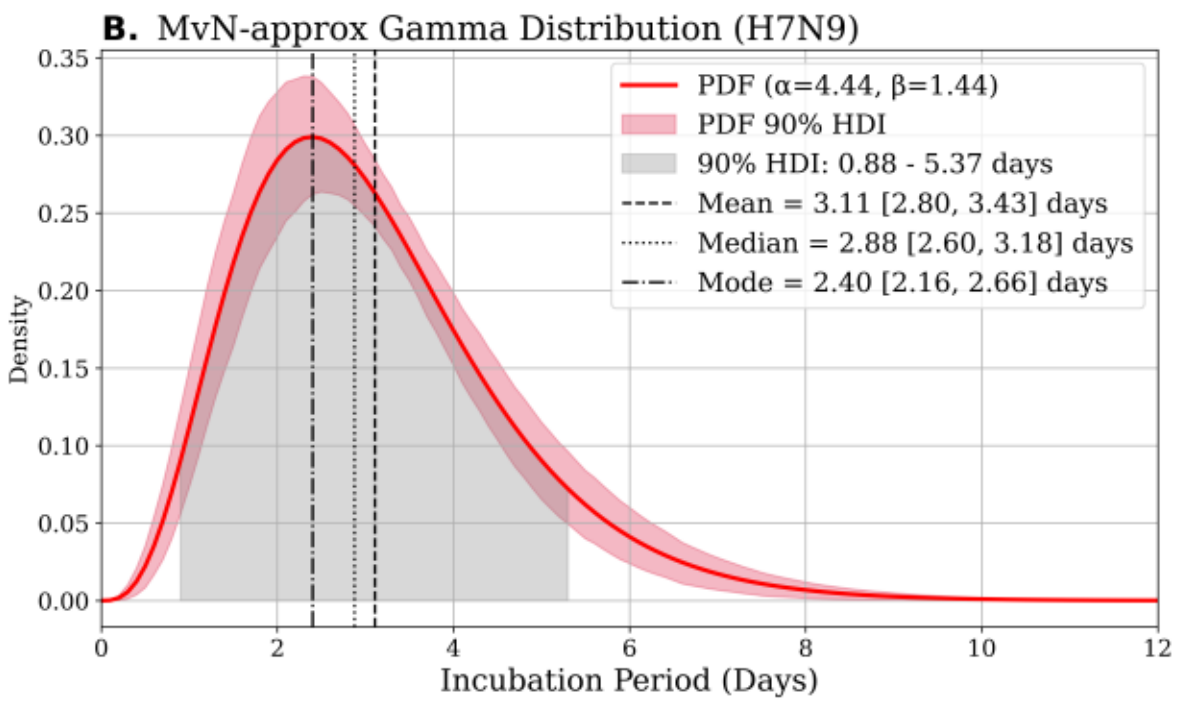

**Figure S7**. H7N9 Analysis PDF summaries. Gamma distributions are computed as $f(x|\alpha,\beta)$ where $f$ is the PDF of the Gamma distribution and $\alpha$ and $\beta$ are shape and rate (their posterior means are shown in plots). The grey shadow indicates the 90% distribution highest density interval (HDI). Dashed, dotted and das-dotted line indicate mean, median and mode of the Gamma distribution respectively, their respective HDIs in brackets.

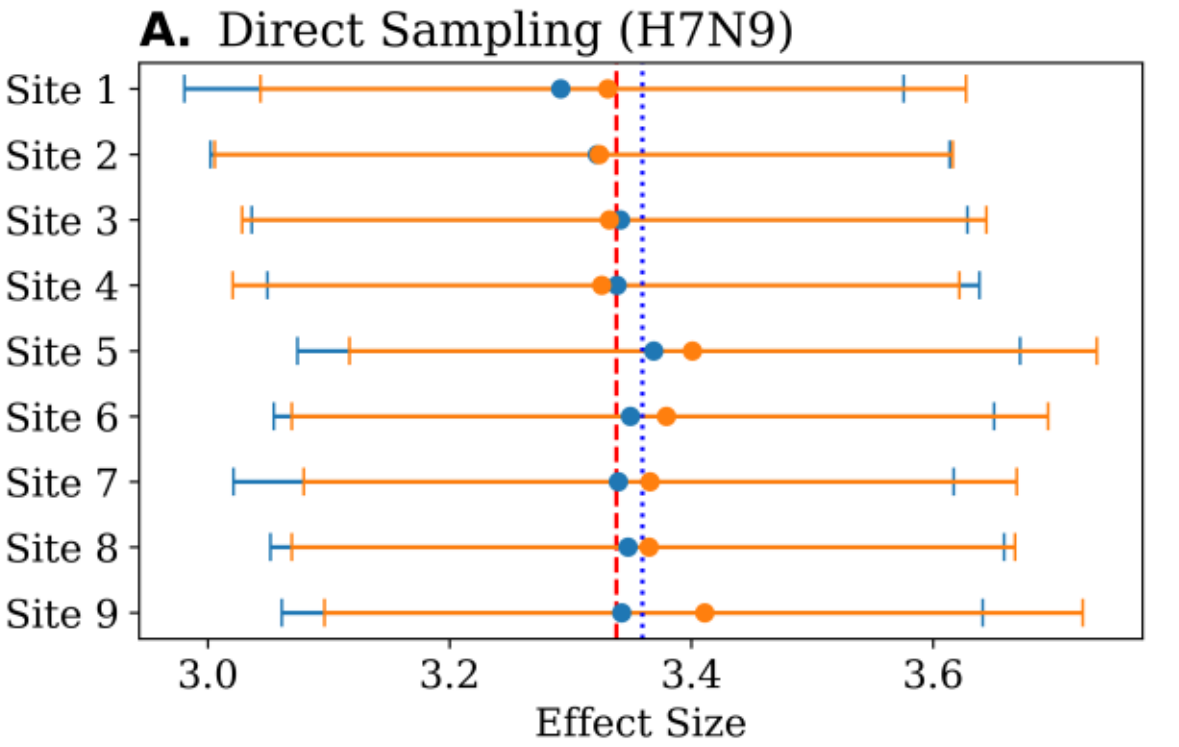

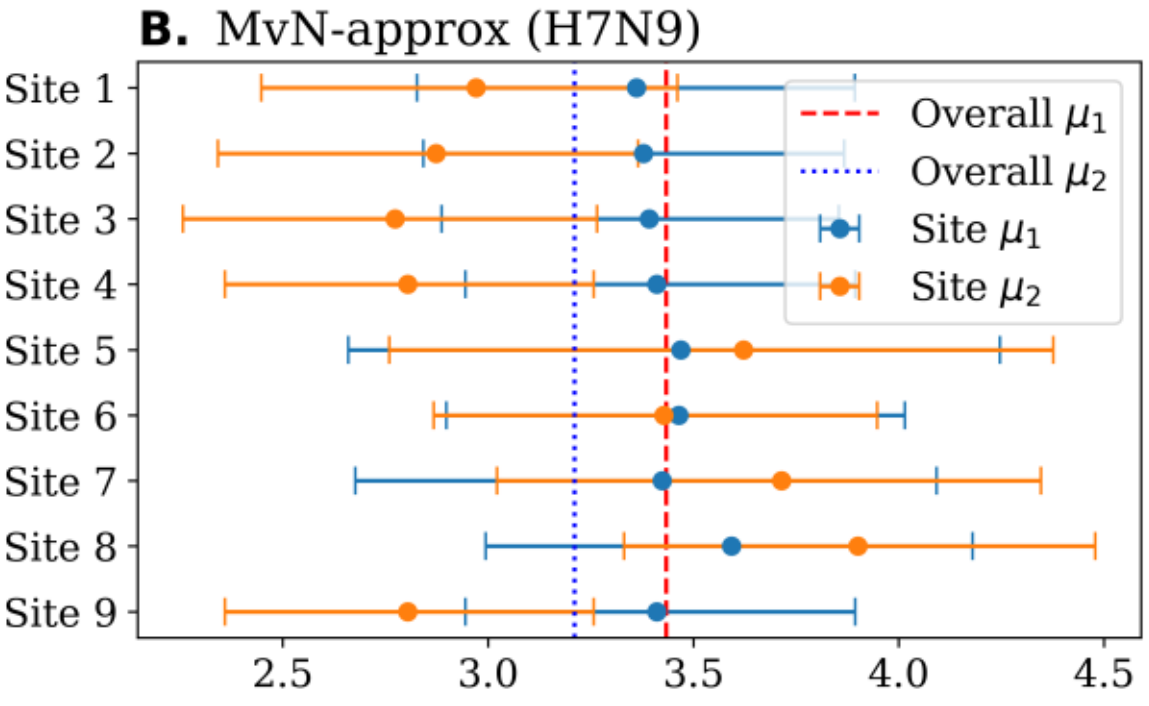

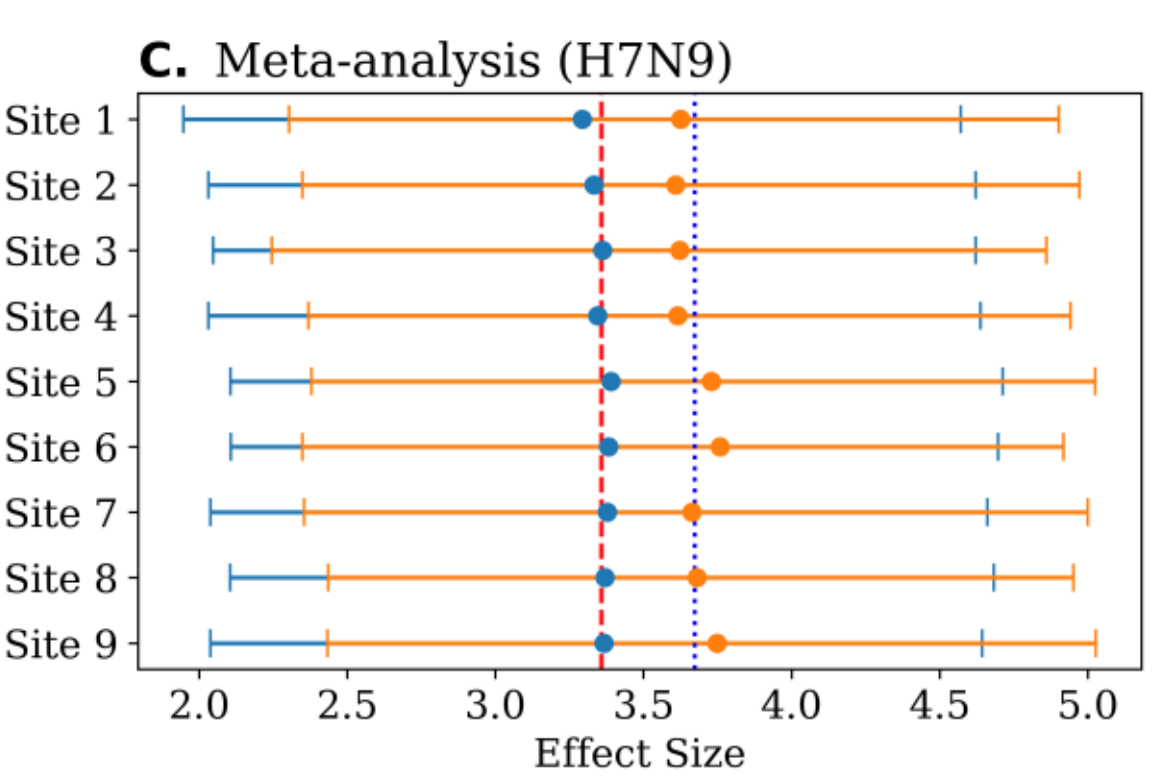

**Figure S8**. H7N9 Analysis Gamma mean ($\mu$) summaries. $\mu_1$ and $\mu_2$ corrspond to the group 1 and group 2 (G1 and G2) from H7N9 models, where G1 are reovered patients and G2 are fatal cases. **A**: Direct Sampling effect-sizes refer to the site means $\mu$ estimated from a whole data hierarchical model (i.e. varying parameters). **B**: MvN-approx. effect-sizes refer to the $\mu$ estimates from a locally sampled fixed model, where models after the firstly smapled receive $\mu_{n-1}$ information to parametrise priors. **B**: Meta-analysis effect-sizes refer to the meta-analysis hierarchical model site-by-site estimated means $\mu$ ($\theta_n$ in the meta-anaysis hierarchical model).

## *S3. Additional COVID-19 Summaries*

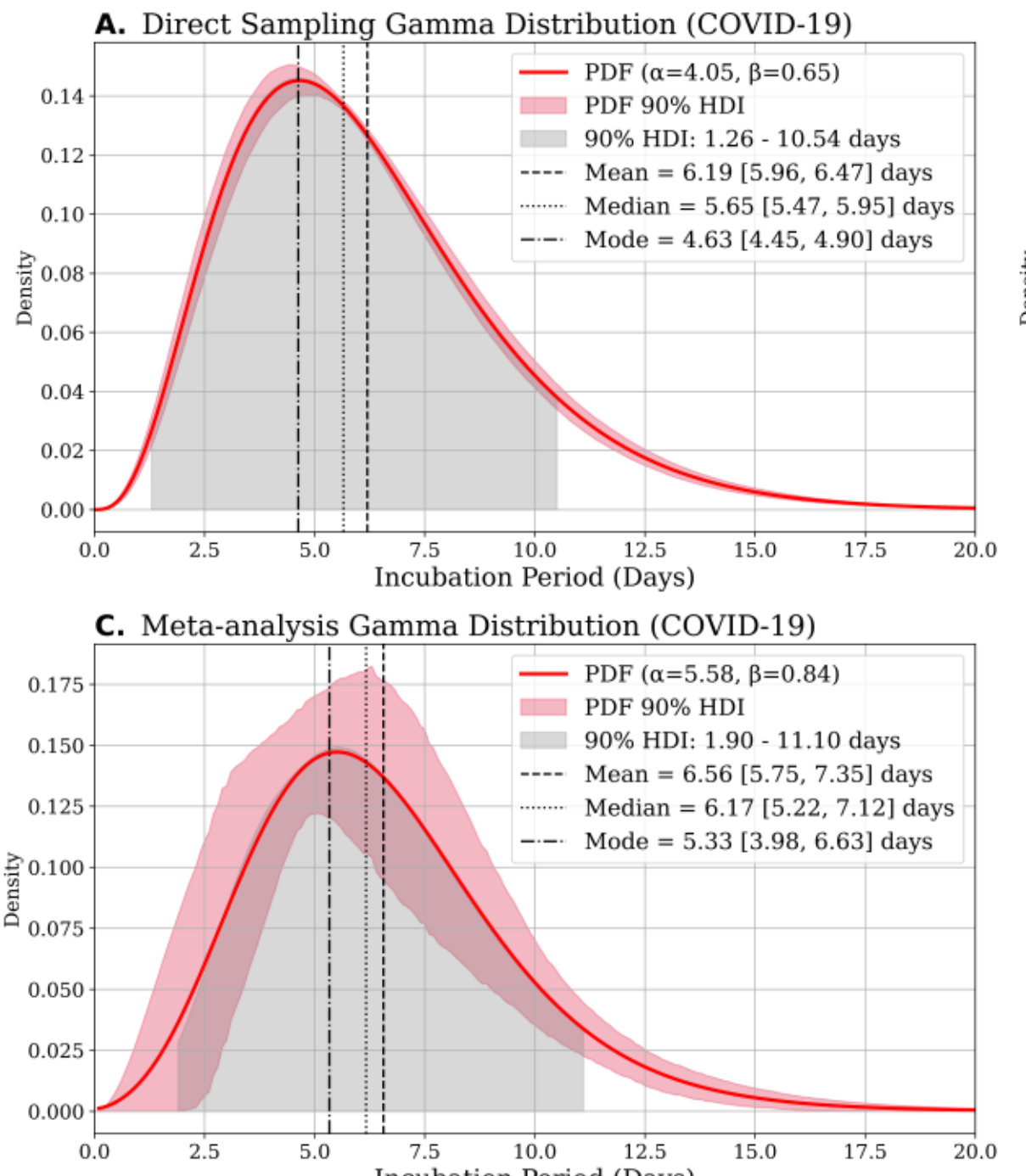

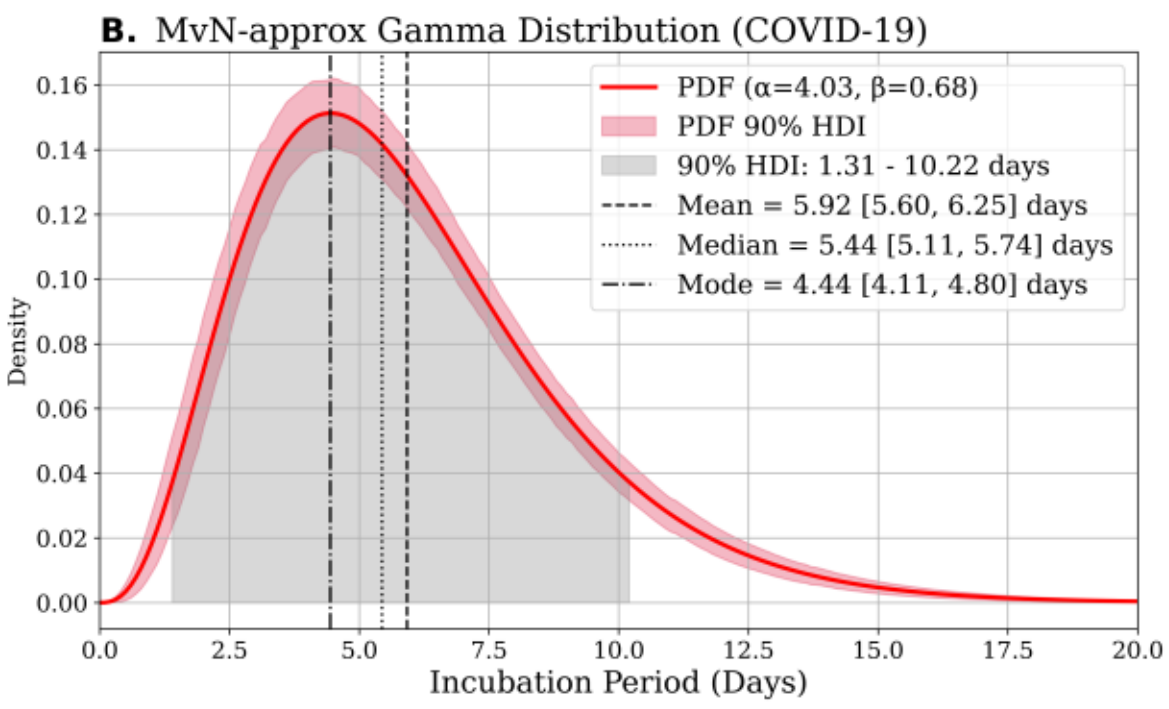

**Figure S9**. COVID-19 Analysis PDF summaries. Gamma distributions are computed as $f(x|\alpha,\beta)$ where $f$ is the PDF of the Gamma distribution and $\alpha$ and $\beta$ are shape and rate (their posterior means are shown in plots). The grey shadow indicates the 90% distribution highest density interval (HDI). Dashed, dotted and das-dotted line indicate mean, median and mode of the Gamma distribution respectively, their respective HDIs in brackets.

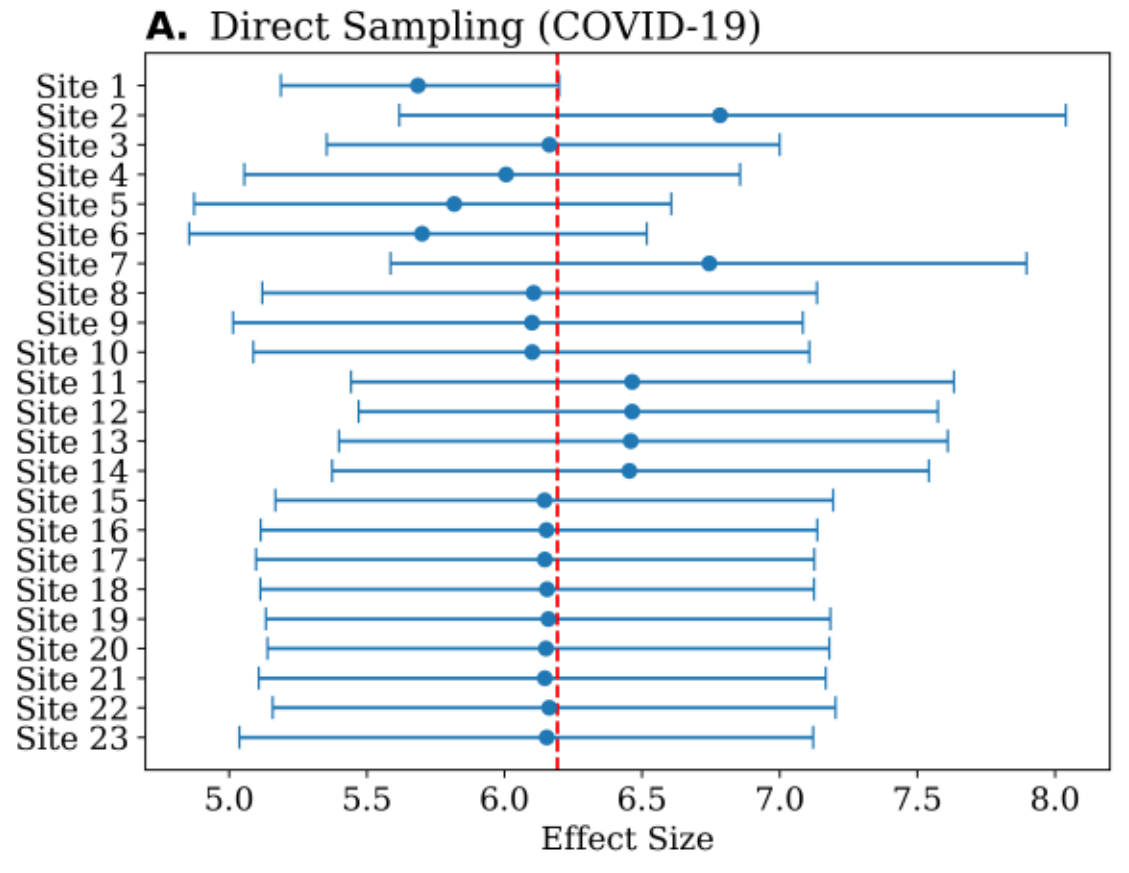

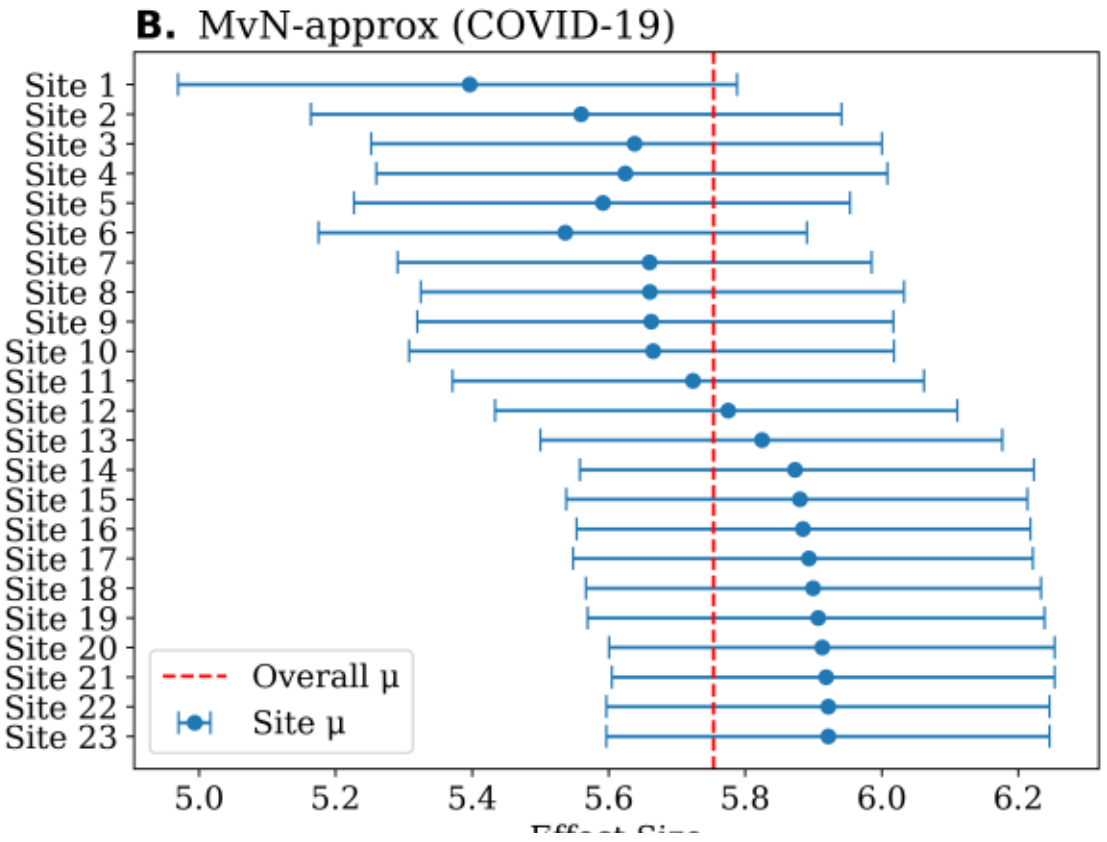

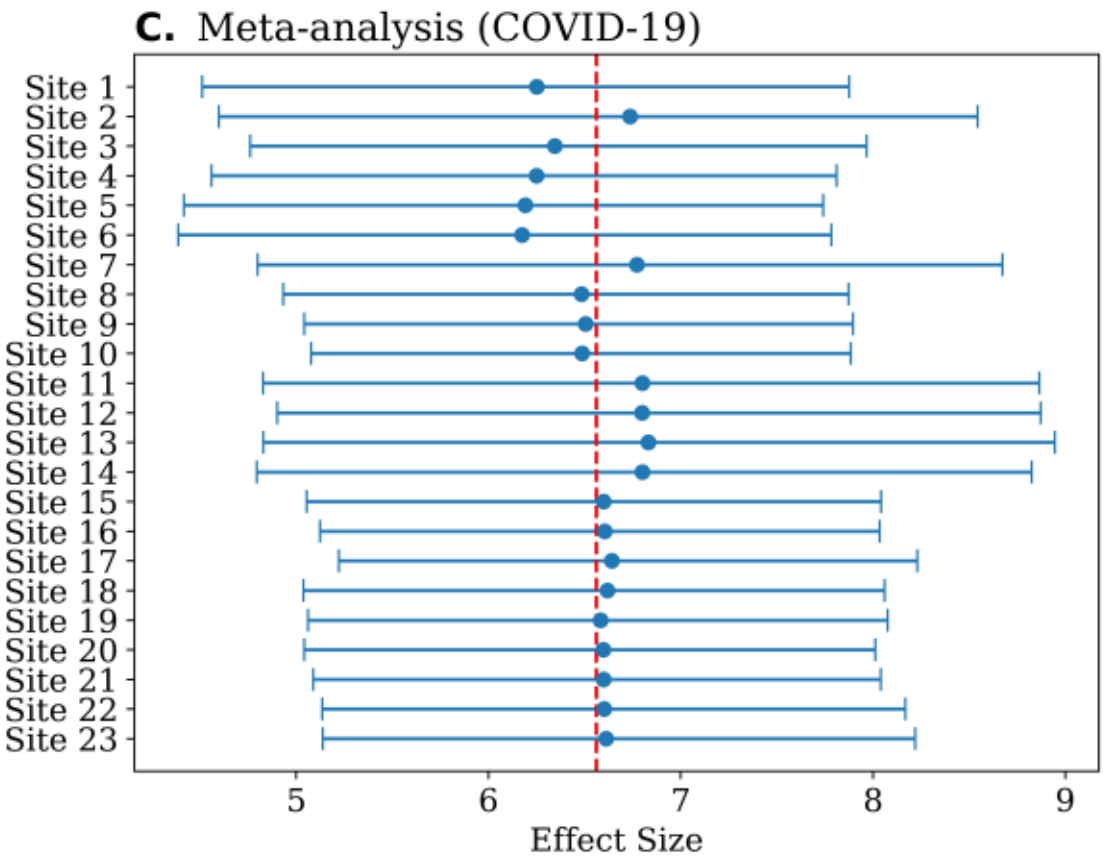

**Figure S10**. COVID-19 Analysis Gamma mean (μ) summaries. **A**: Direct Sampling effect-sizes refer to the site means $\mu$ estimated from a whole data hierarchical model (i.e. varying parameters). **B**: MvN-approx. effect-sizes refer to the $\mu$ estimates from a locally sampled fixed model, where models after the firstly smapled receive $\mu_{n-1}$ information to parametrise priors. Note how sites with very low sample-size [11-23] pull the estimate upwards, despite the strong influence of larger sample-size sites [1-6]. **C**: Meta-analysis effect-sizes refer to the meta-analysis hierarchical model site-by-site estimated means $\mu$ ($\theta_n$ in the meta-anaysis hierarchical model).